\documentclass[10pt,aps,pre, floats, floatfix, superscriptaddress, twocolumn,nofootinbib,amsmath,amssymb,showpacs,showkeys]{revtex4-2}
\usepackage{enumerate}
\usepackage{graphicx}
\usepackage{multirow}
\usepackage{xcolor}
\usepackage{tablefootnote}
\usepackage{threeparttable}
\usepackage{bm}
\usepackage[normalem]{ulem}
\usepackage{comment}
\usepackage{color}
\usepackage{ulem}
\usepackage[hidelinks]{hyperref}

\begin{document}
\title{Inflation without an Inflaton II: observational predictions}

\author{Marisol Traforetti}
\email{marisol.traforetti@icc.ub.edu}
\affiliation{ICCUB, University of Barcelona, Mart\'i i Franqu\`es, 1, E08028 Barcelona, Spain}

\author{Mariam Abdelaziz}
\email{mariamtarekmohamed.abdelaziz-ssm@unina.it}
\affiliation{Scuola Superiore Meridionale, Largo San Marcellino 10, I-80138 Napoli, Italy}
\affiliation{INFN, Sezione di Napoli, Via Cinthia Edificio 6, I-80126 Napoli, Italy}
\affiliation{Department of Astronomy, Space Science and Meteorology, Cairo University,  12613 Giza, Egypt}

\author{Daniele Bertacca}
\email{daniele.bertacca@pd.infn.it}
\affiliation{Dipartimento di Fisica e Astronomia Galileo Galilei, \\Universit\`a degli Studi di Padova, via Marzolo 8, I-35131, Padova, Italy}
\affiliation{INFN, Sezione di Padova, via Marzolo 8, I-35131, Padova, Italy}
\affiliation{INAF- Osservatorio Astronomico di Padova, \\ Vicolo dell Osservatorio 5, I-35122 Padova, Italy}

\author{Raul Jimenez}
\email{raul.jimenez@icc.ub.edu}
\affiliation{ICCUB, University of Barcelona, Mart\'i i Franqu\`es, 1, E08028 Barcelona, Spain}
\affiliation{ICREA, Pg. Lluis Companys 23, Barcelona, 08010, Spain.} 

\author{Sabino Matarrese}
\email{sabino.matarrese@pd.infn.it}
\affiliation{Dipartimento di Fisica e Astronomia Galileo Galilei, \\Universit\`a degli Studi di Padova, via Marzolo 8, I-35131, Padova, Italy}
\affiliation{INFN, Sezione di Padova, via Marzolo 8, I-35131, Padova, Italy}
\affiliation{INAF- Osservatorio Astronomico di Padova, \\ Vicolo dell Osservatorio 5, I-35122 Padova, Italy}
\affiliation{Gran Sasso Science Institute, Viale F. Crispi 7, I-67100 L'Aquila, Italy}

\author{Angelo Ricciardone}
\email{angelo.ricciardone@unipi.it}
\affiliation{Dipartimento di Fisica ``Enrico Fermi'', Universit\`a di Pisa, Pisa I-56127, Italy}
\affiliation{INFN sezione di Pisa, Pisa I-56127, Italy}

\begin{abstract}
We present a complete computation of the scalar power spectrum in the \emph{inflation without inflaton} (IWI) framework, where the inflationary expansion is driven solely by a de Sitter (dS) background and scalar fluctuations arise as second-order effects sourced by tensor perturbations. By explicitly deriving and numerically integrating the full second-order kernel of the Einstein equations, we obtain a scale-invariant scalar spectrum without invoking a fundamental scalar field. In this framework, the amplitude of the scalar fluctuations is directly linked to the scale of inflation.
More precisely, we show that matching the observed level of scalar fluctuations $\Delta_\mathcal{\phi}^2(k_\ast)\approx 10^{-9}$ at Cosmic Microwave Background (CMB) scales fixes the inflationary energy scale $H_{\rm inf}$ as a function of the number of observed e-folds $N_{\rm obs}$. For $N_{\rm obs}\simeq30$–$60$, we find $H_{\mathrm{inf}} \simeq 5 \times 10^{13}\,\mathrm{GeV}$–$2\times10^{10}\,\mathrm{GeV}$, corresponding to a tensor-to-scalar ratio $r \simeq 0.01–5\times10^{-9}$. In particular, requiring consistency with instantaneous reheating, we predict a number of e-foldings $\mathcal{O}(50)$ and an inflationary scale $H_{\mathrm{inf}} \simeq 10^{11}\,\mathrm{GeV}$. We also incorporate in our framework the quantum break-time of the dS state, and we show that it imposes an upper bound on the number of particle species. Specifically, using laboratory constraints on the number of species limits the duration of inflation to $N_{\mathrm{obs}} \lesssim 126$ e-folds.
These results establish the IWI scenario as a predictive and falsifiable alternative to standard inflaton-driven models, linking the observed amplitude of primordial fluctuations directly to the quantum nature and finite lifetime of dS space. 
\end{abstract}

\pacs{98.80.Cq, 98.80.Es, 04.62.+v}
\keywords{cosmology: inflation, primordial perturbations, de Sitter space, quantum gravity, tensor modes}

\maketitle

\section{Introduction}

The inflationary paradigm describes a sufficiently long period of accelerated expansion in the early Universe that naturally solves the hot Big Bang shortcomings, while also providing a mechanism for generating the seeds necessary for the formation of cosmic structures \cite{Starobinsky:1979ty,Starobinsky:1980te,Guth:1980zm,Linde:1981mu,Mukhanov:1981xt,Albrecht:1982wi,Starobinsky:1982ee,Rubakov:1982df,Guth:1982ec,Linde:1983gd,Kofman:1985aw}. As a result, inflation predicts a universe that is homogeneous, isotropic and spatially flat at a background level, with adiabatic, almost Gaussian, and approximately scale-invariant primordial density perturbations. \\
In the simplest inflationary models, these primordial perturbations originate from the amplification of quantum vacuum fluctuations of the inflaton, i.e., the scalar field that drives inflation. 
In addition to these scalar fluctuations, tensor fluctuations (primordial gravitational waves) are also present during inflation.
Both scalar and tensor modes are stretched to super-horizon scales during inflation and subsequently re-enter the horizon as classical density fluctuations and a stochastic background of gravitational waves, respectively.\\
To this day, a large variety of inflationary potentials (i.e., models) can be constructed to match current observations. For this reason, the inflationary paradigm remains intrinsically model-dependent.\footnote{See, e.g., Ref.~ \cite{Martin:2013tda} to get a sense of the size of the inflationary landscape.}
In recent works, attempts have been made to reproduce the principal features of inflation without invoking an inflaton field (see, e.g., Refs. \cite{Gomez:2020xdb,Gomez:2021hbb,Gomez:2021yhd}).
Motivated by the search for a model-independent description of inflationary dynamics, we investigate the predictions of a novel scenario introduced in Ref. \cite{Bertacca:2024zfb}.

In this framework, inflation is driven solely by a dS background, and scalar fluctuations result from second-order effects, sourced by first-order tensor perturbations, which naturally arise from quantum vacuum oscillations in dS. While scalar, vector, and tensor perturbations decouple at linear order, the same does not hold at higher orders in perturbation due to the non-linearity of Einstein’s field equations \cite{10.1143/PTP.45.1747,10.1143/PTP.47.416,Matarrese:1997ay}. Recently, tensor-induced scalar perturbations were studied in Refs. \cite{Bari:2021xvf,Bari:2022grh}.
Additionally, previous works have studied the intrinsic instability of dS spacetime and argued that it provides a graceful exit from inflation without requiring a reheating phase \cite{Mottola:1984ar,Antoniadis:2006wq,Polyakov:1982ug,Dvali:2017eba,Mottola:1985ee,Mottola:1985qt,Alicki:2023rfv,Alicki:2023tfz}. If one describes dS as a quantum coherent state of gravitons, then their self-coupling— as well as their coupling to other relativistic particle species, such as those in the Standard Model, which must always be present—leads to quantum scattering and decay of the same gravitons in dS \cite{Dvali:2017eba}. 
In particular, a higher-order process of particle production occurs, in which the produced particles recoil against all remaining gravitons. 
This allows for the production of low-energy particles.
In this case, the final quantum state can no longer be described as a coherent state, and it departs from the classical mean field evolution.\\

This paper presents a follow-up analysis of Ref. \cite{Bertacca:2024zfb}, aiming to evaluate the power spectrum of scalar fluctuations and provide specific predictions for this mechanism.
The paper is organized as follows. In Section \ref{sec: Second-order Einstein field equations}, we introduce our setup and summarize key results from \cite{Bertacca:2024zfb}. In Section \ref{sec: power spectrum}, we compute the full expression for the power spectrum of scalar fluctuations and show that it exhibits scale-invariance. Then, in Section \ref{sec: The energy scale of inflation}, we discuss the results for the scalar power spectrum. In particular, we show how this framework offers a way to connect the amplitude of scalar fluctuations to the energy scale of inflation $H_{\mathrm{inf}}$ and the number of e-folds $N_{\rm obs}$. We conclude in Section \ref{sec: Conclusions}. Additional details are provided in Appendices \ref{sec:Fourier space conventions}-\ref{sec:The scalar spectral tilt}. 

\section{Second-order Einstein field equations}
\label{sec: Second-order Einstein field equations}

Using the same setup as Ref. \cite{Bertacca:2024zfb}, we work on a pure dS metric and assume Einstein gravity.
We use the following perturbed line element
\begin{eqnarray}
    d s^2 &= & a^2(\eta)\bigg\{ -\left(1 + \psi ^{(2)} \right)d\eta^2  \nonumber \\
    \quad\quad && +\left[\left(1 - \phi ^{(2)}\right)\delta _{ij}  +\chi^{(1)}_{ij} \right]dx^idx^j \bigg\}\,, 
\end{eqnarray}
where $\psi ^{(2)}$ and $\phi ^{(2)}$ denote the second-order scalar perturbations and $\chi^{(1)}_{ij}$ are the first-order tensor fluctuations.

On the right-hand side of Einstein’s field equations, we include the energy–momentum tensor of a generic fluid. As shown in \cite{Bertacca:2024zfb}, such a contribution inevitably arises from the vacuum expectation value of the second-order part of the Einstein tensor due to gravitational waves (GWs). On sub-horizon scales, these GWs generate non-vanishing energy density $\rho$, pressure $p$, and anisotropic stress $\pi_\nu^\mu$. Namely, $T_{\mu\nu}$ reads
\begin{eqnarray}
    T^\mu_\nu=(\rho+p)u^\mu u_\nu+p\delta^\mu_\nu+\pi^\mu_\nu\;,
\end{eqnarray}
where $u^\mu$ is the four-velocity normalized to $u^\mu u_\mu=-1$. 
In addition to this, we shall account for the cosmological constant $\Lambda$ driving the dS expansion.
We consider a generic $w$ and $c_s$, where $w=\Bar{p}/\Bar{\rho}$ is the background equation of state and $c^2_s=p^{(2)}/\rho^{(2)}$ is the adiabatic speed of sound.

The evolution of second-order perturbations is given by Einstein's field equations at second order. In particular, using the traceless part of the $ij$ components and the very large scale limit, we find~\cite{Bertacca:2024zfb} 
\begin{eqnarray}
        \label{eq: phi_second_order}
    \psi_2=\phi_2-8\pi G a^2\Pi_2-\dfrac{{\cal F_\chi}}{4}\;,
\end{eqnarray}
where $\Pi_2$ is the second-order scalar part of the anisotropic stress. In \eqref{eq: phi_second_order}, the quantity $\mathcal{F}_\chi$ is defined in terms of couplings of first-order tensors as
\begin{eqnarray}
     \label{eq:F_chi}
{\cal F_\chi}&=& 
4\nabla^{-2}\bigg(
\frac{3}{4}\chi_1^{lk,m}\chi_{1kl,m} + \dfrac{1}{2} \chi_1^{kl} \nabla^2 \chi_{1lk} - \dfrac{1}{2} \chi_{1,l}^{km}\chi_{1m,k}^l\bigg)\nonumber\\
&&-6 \nabla^{-4}\partial_i\partial^j {\cal A}^i_j\;,
\end{eqnarray}
where
 \begin{eqnarray}\label{eq:calA}
{\cal A}^i_j&=&{1\over     2}\chi_1^{lk,i}\chi_{1kl,j}+\chi_1^{kl}\chi_{1lk,j}^{,i} -\chi_1^{kl}\chi_{1l,jk}^i\nonumber\\
&&-\chi_1^{kl}\chi_{1lj,k}^{,i}+\chi_1^{kl}\chi_{1j,kl}^i+\chi_{1,l}^{ik}\chi_{1jk}^{,l}
-\chi_{1,l}^{ki}\chi_{1j,k}^l\;.\quad
\end{eqnarray}

The equation of motion for the potential $\phi_2$ can be found in Ref.~\cite{Bertacca:2024zfb}.

\section{Power spectrum}
\label{sec: power spectrum}

As a simple analytical case, let us consider the case in which $\psi_2=0$ and $\Pi_2=0$, so we concentrate on the graviton part.
Therefore, from Eq. \eqref{eq: phi_second_order}, the solution in which $\phi_2$ is constant on very large scales reads 
\begin{equation}
\phi_{2}= \frac{1}{4}\cal F_\chi \;.
\end{equation}
The primordial power spectrum of scalar fluctuations $\phi$ is
\begin{eqnarray}
\label{power spectrum definition}
    \langle \phi(\mathbf{k}) \phi(\mathbf{k'}) \rangle &=& \frac{1}{4}  \langle \phi_2(\mathbf{k}) \phi_2(\mathbf{k'}) \rangle \nonumber \\
    &=& (2 \pi)^3 \delta^{(3)} (\mathbf{k+k'}) \mathcal{P}_{\phi} (k)\;.
\end{eqnarray}
The dimensionless scalar power spectrum $\Delta^2_\phi(k)$ is defined as 
\begin{eqnarray}
\label{eq: def dimensionless scalar power spectrum}
    \Delta^2_\phi(k)=\dfrac{k^3}{2\pi^2}\mathcal{P}_\phi(k)\;.
\end{eqnarray}
Since Eq. \eqref{eq:F_chi} contains two terms, we decide to call the first terms in \eqref{eq:F_chi} collectively as `A' and the second term, i.e., $-6\nabla^{-4}\partial_i\partial^j\mathcal{A}^i_j$, as `B'.
Therefore, we can split the contributions to the primordial power spectrum $\mathcal{P}_\phi(k)$ into four parts, namely: $\mathcal{P}^{(AA)}_\phi(k)$, $\mathcal{P}^{(AB)}_\phi(k)$, $\mathcal{P}^{(BA)}_\phi(k)$ and $\mathcal{P}^{(BB)}_\phi(k)$. More precisely, $\mathcal{P}^{(AA)}_\phi(k)$ comes only from the first terms in Eq. \eqref{eq:F_chi}, while $\mathcal{P}^{(BB)}_\phi(k)$ comes from the $\mathcal{A}^i_j$ terms in \eqref{eq:F_chi}. On the other hand, $\mathcal{P}^{(AB)}_\phi(k)$ and $\mathcal{P}^{(BA)}_\phi(k)$ arise from the cross terms between the `A' and `B' contributions and, as we will see, they have the same expression. 
Hence, the full power spectrum is given simply by 
\begin{eqnarray}
\mathcal{P}_\phi(k)=\mathcal{P}^{(AA)}_\phi(k)+\mathcal{P}^{(BB)}_\phi(k)+2\mathcal{P}^{(AB)}_\phi(k)\;. 
\label{eq: full power spectrum}
\end{eqnarray} 
We include details on our Fourier conventions and the power spectrum calculations in Appendices \ref{sec:Fourier space conventions} and \ref{sec:Power Spectrum computation}, respectively.

The first part in \eqref{eq: full power spectrum} was already computed in Ref. \cite{Bertacca:2024zfb}, and we check that it is given by
\begin{eqnarray}
        \mathcal{P}^{(AA)}_\phi(k)&=&\dfrac{1}{64(2\pi)^3}\dfrac{1}{k^4}\int d^3k_1d^3k_2\,\delta^{(3)}(\mathbf{k}-(\mathbf{k}_1+\mathbf{k}_2))\nonumber\\
&&\times\mathcal{K}^{(AA)}_h(\mathbf{k}_1,\mathbf{k}_2,k^2)\mathcal{P}_h(k_1)\mathcal{P}_h(k_2)\;,
    \label{PS first part- first part}
\end{eqnarray}
where $\mathcal{P}_h(k_i)$ is the tensor power spectrum on super-horizon scales in dS and it is defined as
\begin{eqnarray}
\label{tensor power spectrum dS}
    \mathcal{P}_h (k) = \frac{2 \pi^2}{k^3}\Delta_h^2(k)=\frac{2 \pi^2}{k^3} \left[\frac{16}{\pi} \left (  \frac{H_{\rm inf}}{m_{\rm pl}}   \right )^2\right]\;,
\end{eqnarray}  
with $\Delta_h^2(k)$ being the dimensionless primordial tensor power spectrum, $H_{\mathrm{inf}}=\sqrt{\Lambda/3}$ is the Hubble constant during inflation and $m_{\rm pl}=G^{-1/2}$ is the Planck mass. Eq. \eqref{tensor power spectrum dS} already accounts for the two independent polarization states with the same power spectrum.
In Eq. \eqref{PS first part- first part}, we also defined the kernel $\mathcal{K}^{(AA)}_h(\mathbf{k}_1,\mathbf{k}_2,k^2)$ as 
\begin{widetext}
\begin{eqnarray}
\mathcal{K}^{(AA)}_h(\mathbf{k}_1,\mathbf{k}_2,k^2)&=&\dfrac{1}{4}\left\{ (k^2_1+k^2_2+3{\mathbf{k}}_1\cdot {\mathbf{k}}_2)^2\left[(1-\hat{\mathbf{k}}_1\cdot \hat{\mathbf{k}}_2)^4+(1+\hat{\mathbf{k}}_1\cdot \hat{\mathbf{k}}_2)^4\right]\right.\nonumber\\
        &&\left.\quad+8k^2_1k^2_2(\hat{\mathbf{k}}_1\times \hat{\mathbf{k}}_2)^4\left[1+(\hat{\mathbf{k}}_1\cdot \hat{\mathbf{k}}_2)^2\right]+8(\mathbf{k}_1\cdot\mathbf{k}_2)(k^2_1+k^2_2+3{\mathbf{k}}_1\cdot {\mathbf{k}}_2)(\hat{\mathbf{k}}_1\times \hat{\mathbf{k}}_2)^2\left[3+(\hat{\mathbf{k}}_1\cdot \hat{\mathbf{k}}_2)^2\right] \right\}\,,\quad\quad
    \label{eq: kernel first part - first part}
    \end{eqnarray}
\end{widetext}
where the factor $1/4$ comes from the two tensor power spectra, $\mathcal{P}_h(k_1)$ and $\mathcal{P}_h(k_2)$, as Eq. \eqref{tensor power spectrum dS} accounts for the two independent polarizations.
To this, we add the computation of the primordial power spectrum given by the $\mathcal{A}^i_j$ terms in Eq. \eqref{eq:F_chi} (denoted as `B'), with $\mathcal{A}^i_j$ given in Eq. \eqref{eq:calA}. The expression for $\mathcal{P}^{(BB)}_\phi(k)$ is the following
\begin{eqnarray}
        \mathcal{P}^{(BB)}_\phi(k)&=&\dfrac{9}{64(2\pi)^3}\dfrac{1}{k^8}\int d^3k_1d^3k_2\,\delta^{(3)}(\mathbf{k}-(\mathbf{k}_1+\mathbf{k}_2))\nonumber\\
&&\times\mathcal{K}^{(BB)}_h(\mathbf{k}_1,\mathbf{k}_2,k^2)\mathcal{P}_h(k_1)\mathcal{P}_h(k_2)\;,
    \label{PS second part- second part}
\end{eqnarray}
where the kernel $\mathcal{K}^{(BB)}_h(\mathbf{k}_1,\mathbf{k}_2,k^2)$ is given by
\begin{widetext}
    \begin{eqnarray}        \mathcal{K}^{(BB)}_h(\mathbf{k}_1,\mathbf{k}_2,k^2)&= &\dfrac{1}{4}\left\{ \left[3(k^2_1+k^2_2)\mathbf{k}_1\cdot \mathbf{k}_2+k^2_1k^2_2\left(1+3(\hat{\mathbf{k}}_1\cdot \hat{\mathbf{k}}_2)^2\right)+k^4_1+k^4_2\right]^2\left[(1-\hat{\mathbf{k}}_1\cdot \hat{\mathbf{k}}_2)^4+(1+\hat{\mathbf{k}}_1\cdot \hat{\mathbf{k}}_2)^4\right]
        \right.\nonumber\\
        &&\left.\quad+8\left[3(k^2_1+k^2_2)\mathbf{k}_1\cdot \mathbf{k}_2+k^2_1k^2_2\left(1+3(\hat{\mathbf{k}}_1\cdot \hat{\mathbf{k}}_2)^2\right)+k^4_1+k^4_2\right](k^2_1+k^2_2+{\mathbf{k}}_1\cdot {\mathbf{k}}_2)\right.\nonumber\\
        &&\left.\quad\quad\quad(\mathbf{k}_1\cdot\mathbf{k}_2)(\hat{\mathbf{k}}_1\times \hat{\mathbf{k}}_2)^2\left[3+(\hat{\mathbf{k}}_1\cdot \hat{\mathbf{k}}_2)^2\right]\right.\nonumber\\
        &&\left.\quad+8(k^2_1+k^2_2+{\mathbf{k}}_1\cdot {\mathbf{k}}_2)^2 k^2_1k^2_2(\hat{\mathbf{k}}_1\times \hat{\mathbf{k}}_2)^4\left[1+(\hat{\mathbf{k}}_1\cdot \hat{\mathbf{k}}_2)^2\right] \right\}\;.
        \label{eq: kernel second part - second part}
    \end{eqnarray}
\end{widetext}
Moreover, we need to compute the cross terms between the `A' and `B' terms in Eq. \eqref{eq:F_chi}. The same expressions will hold for both $\mathcal{P}^{(AB)}_\phi(k)$ and $\mathcal{P}^{(BA)}_\phi(k)$, and the kernels $\mathcal{K}^{(AB)}_h(\mathbf{k}_1,\mathbf{k}_2,k^2)$ and $\mathcal{K}^{(BA)}_h(\mathbf{k}_1,\mathbf{k}_2,k^2)$.
\\
\\
\\
Namely, for $\mathcal{P}^{(AB)}_\phi(k)$ we get
\begin{eqnarray}
        \mathcal{P}^{(AB)}_\phi(k)&=&-\dfrac{3}{64(2\pi)^3}\dfrac{1}{k^6}\int d^3k_1d^3k_2\,\delta^{(3)}(\mathbf{k}-(\mathbf{k}_1+\mathbf{k}_2))\nonumber\\
&&\times\mathcal{K}^{(AB)}_h(\mathbf{k}_1,\mathbf{k}_2,k^2)\mathcal{P}_h(k_1)\mathcal{P}_h(k_2)\;,
    \label{PS first part- second part}
\end{eqnarray}
where the kernel $\mathcal{K}^{(AB)}_h(\mathbf{k}_1,\mathbf{k}_2,k^2)$ is
\begin{widetext}
\begin{eqnarray}  
        \mathcal{K}^{(AB)}_h(\mathbf{k}_1,\mathbf{k}_2,k^2)&=
            &\dfrac{1}{4}\left\{ (k^2_1+k^2_2+3{\mathbf{k}}_1\cdot {\mathbf{k}}_2)\left[3(k^2_1+k^2_2)\mathbf{k}_1\cdot \mathbf{k}_2+k^2_1k^2_2\left(1+3(\hat{\mathbf{k}}_1\cdot \hat{\mathbf{k}}_2)^2\right)+k^4_1+k^4_2\right]\right.\nonumber\\
        &&\left.\quad\left[(1-\hat{\mathbf{k}}_1\cdot \hat{\mathbf{k}}_2)^4+(1+\hat{\mathbf{k}}_1\cdot \hat{\mathbf{k}}_2)^4\right]
        +4(\mathbf{k}_1\cdot\mathbf{k}_2)(\hat{\mathbf{k}}_1\times \hat{\mathbf{k}}_2)^2\left[3+(\hat{\mathbf{k}}_1\cdot \hat{\mathbf{k}}_2)^2\right]\right.\nonumber\\
        &&\left.\quad\left[(k^2_1+k^2_2+3{\mathbf{k}}_1\cdot {\mathbf{k}}_2)(k^2_1+k^2_2+{\mathbf{k}}_1\cdot {\mathbf{k}}_2)+\left[3(k^2_1+k^2_2)\mathbf{k}_1\cdot \mathbf{k}_2+k^2_1k^2_2\left(1+3(\hat{\mathbf{k}}_1\cdot \hat{\mathbf{k}}_2)^2\right)+k^4_1+k^4_2\right]\right] \right.\nonumber\\
&&\left.\quad+8(k^2_1+k^2_2+\mathbf{k}_1\cdot\mathbf{k}_2)k^2_1k^2_2(\hat{\mathbf{k}}_1\times \hat{\mathbf{k}}_2)^4\left[1+(\hat{\mathbf{k}}_1\cdot \hat{\mathbf{k}}_2)^2\right]\right\}\;.
        \label{eq: kernel first part - second part}
        \end{eqnarray}
\end{widetext}
As already noticed in Ref. \cite{Bertacca:2024zfb}, qualitatively, we see from Eqs. \eqref{PS first part- first part}, \eqref{PS second part- second part}, and \eqref{PS first part- second part} that we expect the dimensionless power spectrum \eqref{eq: def dimensionless scalar power spectrum} to be scale invariant as no uncompensated powers of the momenta appear from the expressions of the kernels in Eqs. \eqref{eq: kernel first part - first part}, \eqref{eq: kernel second part - second part} and \eqref{eq: kernel first part - second part}, respectively. Formally, it can be shown that there is no preferred scale since $\Delta^2(k)$ is left unchanged when we perform a global rescaling of the momenta as $k\rightarrow\lambda k$.\\

To solve the integrals in Eqs. \eqref{PS first part- first part}, \eqref{PS second part- second part} and \eqref{PS first part- second part}, we first use $\delta^{(3)}(\mathbf{k}-(\mathbf{k}_1+\mathbf{k}_2))$ to perform the integral over $d^3k_2$, so $\mathbf{k}_2=\mathbf{k}-\mathbf{k}_1$. Due to isotropy, we are free to align $\mathbf{k}$ with the z-axis, i.e., $\mathbf{k}=k(0,0,1)$, and we write $\mathbf{k}_1$ in spherical coordinates as $$\mathbf{k}_1=k_1(\sin\theta_1\cos\varphi_1,\sin\theta_1\sin\varphi_1,\cos\theta_1)\;,$$ where $\theta_1\in[0,\pi]$ and $\varphi_1\in[0,2\pi]$. Then, it follows that $$\mathbf{k}-\mathbf{k}_1=(-k_1\sin\theta_1\cos\varphi_1,-k_1\sin\theta_1\sin\varphi_1,k-k_1\cos\theta_1)\;.$$ The integral over the wave-number $\mathbf{k}_1$ is $$\int d^3k_1=\int^\infty_0dk_1k_1^2\int^\pi_0d\theta_1\sin{\theta}_1\int^{2\pi}_0d\varphi_1\;,$$ where the integration over $\varphi_1$ is trivial and it gives a factor $2\pi$.
Similar to what is usually done for scalar-induced GWs \cite{Kohri:2018awv,Pi:2020otn,Perna:2024ehx}, it is useful to work in terms of the dimensionless variables 
\begin{eqnarray}
x = \dfrac{k_1}{k} \quad\text{and}\quad y =\dfrac{k_2}{k}= \dfrac{|\mathbf{k}-\mathbf{k}_1|}{k}\;. 
\label{eq: def dimensionless variables x and y}
\end{eqnarray}
The Jacobian of the transformation is $$J=k\dfrac{y}{x}\dfrac{1}{\sin{\theta_1}(x,y)}\;.$$
In this way, the integral over the wave-number $\mathbf{k}_1$ becomes 
$$\int d^3k_1\rightarrow(2\pi)k^3\int^\infty_0dx\int^{x+1}_{|x-1|}dy (xy)\;.$$
Moreover, using the cosine theorem we have that $$|\mathbf{k}-\mathbf{k}_1|^2=k^2+k_1^2-2k\,k_1\cos\theta_1\;,$$ and we can write $\cos\theta_1$ in terms of the variables $\{x,y\}$ defined in \eqref{eq: def dimensionless variables x and y} as $$\cos\theta_1=\dfrac{1+x^2-y^2}{2x}\;.$$ Next, through an explicit calculation, we evaluate the scalar and vector products between $\hat{\mathbf{k}}_1$ and $\hat{\mathbf{k}}_2$, and we obtain 
\begin{eqnarray}
\hat{\mathbf{k}}_1\cdot\hat{\mathbf{k}}_2=\dfrac{-x^2+x\cos\theta_1}{xy}
\label{eq: scalar product in terms of x,y}
\end{eqnarray}
and
\begin{eqnarray}
|\hat{\mathbf{k}}_1\times\hat{\mathbf{k}}_2|=\dfrac{\sin\theta_1}{y}=\dfrac{\sqrt{1-\cos^2\theta_1}}{y}\;,
\label{eq: vector product in terms of x,y}
\end{eqnarray}
respectively.\\

Using this setup, we can now evaluate the integrals in Eqs. \eqref{PS first part- first part}, \eqref{PS second part- second part} and \eqref{PS first part- second part} numerically for the input tensor power spectrum given in Eq. \eqref{tensor power spectrum dS}. Then, summing all three contributions, the full scalar power spectrum is given by Eq. \eqref{eq: full power spectrum}. Finally, the dimensionless scalar power spectrum $\Delta^2_\phi(k)$ defined in Eq. \eqref{eq: def dimensionless scalar power spectrum} results
\begin{eqnarray}
    \Delta^2_\phi(k)=\int^\infty_0dx\int^{x+1}_{|x-1|}dy\dfrac{1}{x^2y^2}\mathcal{K}_h(x,y)\left(\dfrac{H_{\mathrm{inf}}}{m_{\rm pl}}\right)^4\;,\quad\quad
    \label{eq: dimensionless power spectrum to compute}
\end{eqnarray}
and, as anticipated before, it turns out to be independent of $k$ as no uncompensated powers of $k$ appeared in $\mathcal{P}_\phi^{(AA)}(k)$, $\mathcal{P}_\phi^{(BB)}(k)$ and $\mathcal{P}_\phi^{(AB)}(k)$. Here, $\mathcal{K}_h(x,y)$ denotes the full kernel and it includes the contributions $\mathcal{K}_h^{(AA)}(x,y)$, $\mathcal{K}_h^{(BB)}(x,y)$ and $\mathcal{K}_h^{(AB)}(x,y)=\mathcal{K}_h^{(BA)}(x,y)$, together with the correct numerical factors. 
For completeness, the expressions for all the kernels in the $\{x,y\}$ variables are reported in Appendix \ref{sec: Full kernel}. Notice that the value of the power spectrum $\Delta^2_\phi(k)$ in Eq. \eqref{eq: dimensionless power spectrum to compute} is given in terms of $(H_{\mathrm{inf}}/m_{\rm pl})^4$. Thus, once we evaluate numerically the integrals in Eq. \eqref{eq: dimensionless power spectrum to compute}, we can directly connect the amplitude of the scalar fluctuations to the energy scale of inflation. 

However, let us notice the following. In the calculations, we assumed that the tensor modes are constant in time, which is valid only if their momenta are $k_1,k_2\ll k_{\rm end}\sim|\eta^{-1}_{\rm end}|$, where $k_{\rm end}$ is the last mode that exited the horizon at the end of inflation and $\eta_{\rm end}$ the conformal time at the end of inflation. Moreover, given an external wave-number $k$, we have solved the equation of motion for the potential $\phi_2$ for very large scales, thus for $k\ll k_{\rm end}\sim|\eta^{-1}_{\rm end}|$. 
Once we fix $k_{\rm end}$ (which is related to the super-horizon requirement), the tensor modes will be outside the horizon for $k_1,k_2<k_{\rm end}$ and inside the horizon for $k_1,k_2>k_{\rm end}$. 

We can relate the maximum value of the wavenumber of the mode leaving the horizon at the end of inflation $k_{\rm end}$ to the number of e-folds $N_{\mathrm{tot}}\equiv\ln{(a_{\mathrm{end}}/a_{\mathrm{in}})}$, from the beginning to the end of inflation. Since the largest cosmological scale that we can probe today is the cosmological horizon\footnote{Notice that we normalized the scale factor at the present time, so $a_0=1$.} $k_0=H_0=10^{-42}\,\rm{ GeV}\simeq1/4000\, \rm{Mpc^{-1}}$, we define the minimum (or observable) number of e-folds $N_{\rm obs}$ as the time from which $k_0$ crosses the horizon during inflation to the end of inflation. In this way, one can show that\footnote{To be more explicit, since $H_{\mathrm{inf}}$ is constant in dS, one can relate $k_{\rm end}$ to $k_\mathrm{in}$, as $$k_{\mathrm{end}}=a_{\mathrm{end}}H_{\mathrm{inf}}=\dfrac{a_\mathrm{end}}{a_\mathrm{in}}a_\mathrm{in}H_\mathrm{inf}=\dfrac{a_\mathrm{end}}{a_\mathrm{in}}k_\mathrm{in}=e^{N_{\rm tot}}k_{\mathrm{in}}\,,$$ where in the last equality we used $N_{\rm tot}\equiv\ln{(a_{\mathrm{end}}/a_{\mathrm{in}})}$. If we take $k_{\mathrm{in}}$ to be the largest observable scale $k_0=H_0$, we are considering the minimum number of e-folds relevant for the observable Universe, i.e. $N_{\mathrm{obs}}\equiv\ln{(a_{\mathrm{end}}/a_0)}$ with $a_0=1$.} $k_{\mathrm{end}}= e^{N_{\mathrm{obs}}} H_0$. 
Therefore, the super-horizon requirement is related to the number of e-folds relevant for the observable Universe. For example, if we take $N_{\mathrm{obs}}=60$ as a reference value, we get $k_{\rm end}\sim10^{22}\,\rm Mpc^{-1}$. \\
This implies that, when we numerically integrate Eq. \eqref{eq: dimensionless power spectrum to compute}, we should stop the integration at a value $x_{\rm max}=~(k_1/k)_{\rm max}= e^{N_{\rm obs}}$ up to which our formula is valid. Notice that here we are considering comoving wavenumbers.

Recall that the relation to the physical wave-number $k_{\rm phys}$ is $k_{\rm phys}=a(\eta)^{-1}k$, where the scale factor in dS is defined as $a(\eta)=-1/(H_{\mathrm{inf}}\,\eta)$ with $-\infty<\eta<0$. \\

\section{Results and Discussion}
\label{sec: The energy scale of inflation}
We discuss here the results for the dimensionless power spectrum of scalar perturbations and for the value of the energy scale of inflation obtained in our framework.

In the previous Section, we have outlined how to obtain a complete expression for the dimensionless scalar power spectrum $\Delta^2_\phi$, which is ultimately given by Eq. \eqref{eq: dimensionless power spectrum to compute}, when filtering out scales smaller than the horizon at the end of inflation.
This means that, once we fix the number of e-foldings $N_{\mathrm{obs}}$ that inflation lasts, the result of the numerical integration will give the scalar power spectrum in units of $(H_{\rm inf}/m_{\rm pl})^4$.
Therefore, notice that in this scenario we have two parameters: the energy scale of de Sitter $H_{\mathrm{inf}}$, and the minimum number of e-folds $N_{\mathrm{obs}}$.\\
Since $\Delta^2_\phi(k_\ast)$ is fixed by observations at the pivot scale $k_\ast$, we can directly obtain the energy scale of inflation $H_{\rm inf}$, given a value of $N_{\rm obs}$.
In turn, the value of $H_{\mathrm{inf}}$ fixes the amplitude of the tensor power spectrum using Eq. \eqref{tensor power spectrum dS}. 
From this, we can compute the tensor-to-scalar ratio, which describes the amplitude of tensor perturbations relative to that one of scalar perturbations as $$r(k_\ast)=\dfrac{\Delta^2_h(k_\ast)}{\Delta^2_\phi(k_\ast)}\,,$$ and it is evaluated at $k_\ast$. 
The latest constraints from \textit{Planck} 2018 gave $\Delta^2_\phi(k_\ast)\simeq2.1\times 10^{-9}$ (TT, TE, EE+low-E+lensing \cite{Planck:2018jri,Planck:2018vyg}). 
Combining data also from  BICEP2/Keck Array and LIGO\&Virgo2016 leads to a limit on the tensor-to-scalar ratio of 
$r_{0.01}<0.066$ at $95\%$ confidence level (CL) \cite{Planck:2018jri,Galloni:2022mok,Tristram:2021tvh}, where the pivot scale is $k_\ast=0.01\;\rm Mpc^{-1}$.\\

\begin{figure}[t!]
    \centering  
    \includegraphics[width=0.49\textwidth]{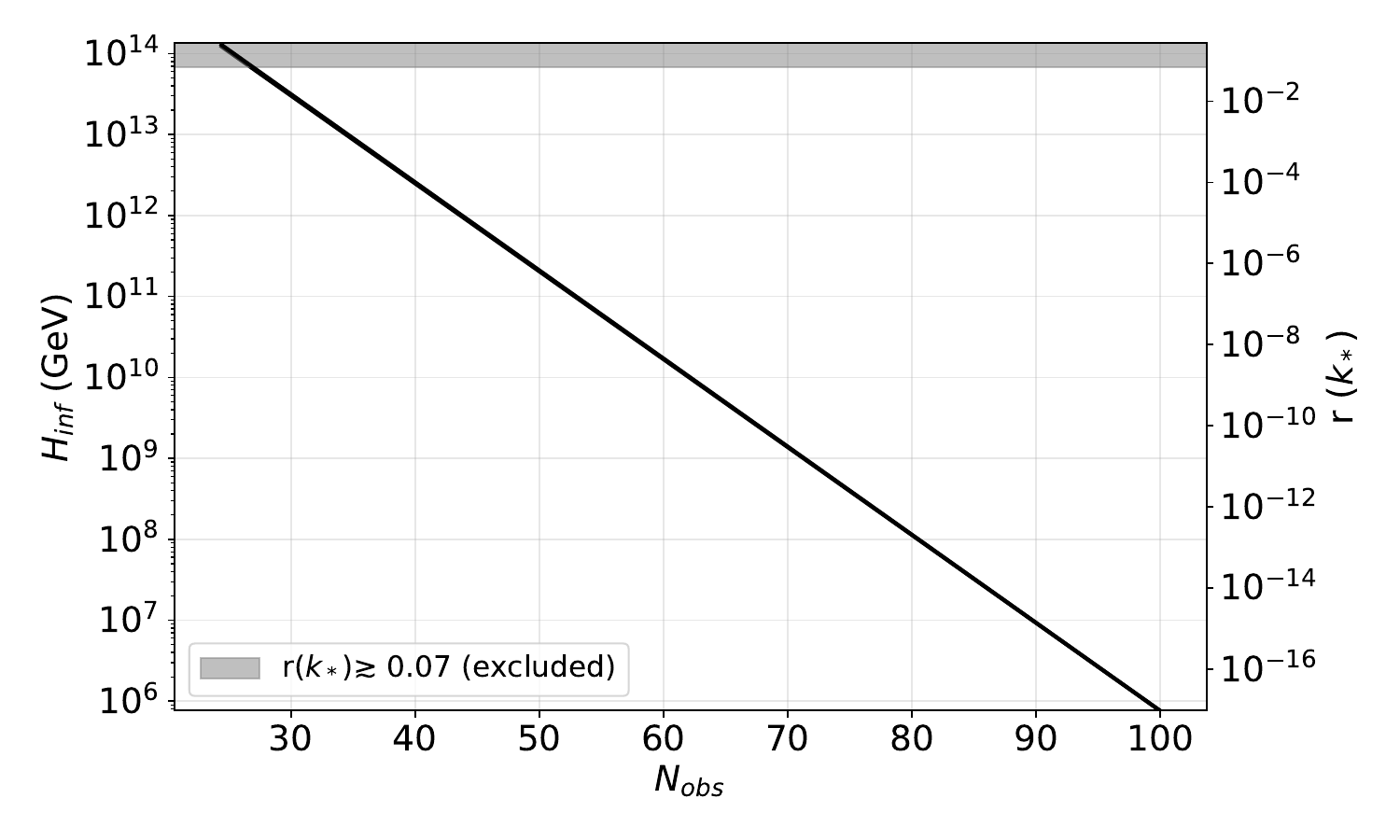}
    \caption{Energy scale of inflation $H_{\rm inf}(\rm GeV)$ and the tensor-to-scalar ratio $r(k_\ast)$ as a function of $N_{\rm obs}$ used in the numerical integration of Eq. \eqref{eq: dimensionless power spectrum to compute}, when implementing the super-horizon requirement and stopping the integration at $x_{\rm max}= e^{N_{\rm obs}}$ as explained in the main text. We show as a gray shaded area the region excluded since it gives a tensor-to-scalar ratio inconsistent with current upper bounds at the CMB pivot scale $k_\ast$, i.e. $r(k_\ast)\gtrsim0.07$ \cite{Planck:2018jri}.}  
    \label{fig: Hinf VS Nmin}  
\end{figure}

Now, we numerically integrate Eq. \eqref{eq: dimensionless power spectrum to compute} for different values of $N_{\rm obs}$, and we fix $H_{\mathrm{inf}}$ in order to match the amplitude of the scalar power spectrum constrained by \textit{Planck} \cite{Planck:2018jri,Planck:2018vyg}. Our pivot scale is set to $k_\ast=~0.01\;\rm Mpc^{-1}$.
In particular, we notice that the main contribution to the numerical integration comes from the kernels $\mathcal{K}^{(AA)}_h(x,y)$ and $\mathcal{K}^{(BB)}_h(x,y)$, while the cross-terms $\mathcal{K}^{(AB)}_h(x,y)$ and $\mathcal{K}^{(BA)}_h(x,y)$ are negligible (see Appendix \ref{sec: Full kernel} for their explicit expressions). In Figure \ref{fig: Hinf VS Nmin}, we plot the energy scale of inflation $H_{\rm inf}$ and the tensor-to-scalar ratio $r$ as a function of $N_{\rm obs}$.

\noindent In the plot, we vary $N_{\mathrm{obs}}$ from 25 to 100, which is equivalent to the range of $k_{\rm end}\in[10^{7},10^{40}]\;\rm Mpc^{-1}$. Notice that the higher the number of e-folds $N_{\rm obs}$, the smaller the scale of inflation we need to match $\textit{Planck}$ constraints of $\Delta_\phi^2(k_\ast)$, thereby reducing the tensor-to-scalar ratio $r(k_\ast)$. Therefore, in this framework, we can obtain the $H_{\mathrm{inf}}$ and $r(k_\ast)$, depending on how much we let inflation last, which is something that our calculations cannot tell us since we have only one observable, $\Delta^2_\phi(k_\ast)$.\\
In particular, for $N_{\mathrm{obs}}=30$, i.e., for $k_{\rm end}\sim10^{9}\;\rm Mpc^{-1}$, we obtain a scale of inflation of the order $H_{\rm inf}\simeq3\times10^{13}\,\rm{GeV}$, which results in a tensor power spectrum\footnote{Recall that the dimensionless tensor power spectrum $\Delta_h^2(k_\ast)$ in dS is defined in Eq. \eqref{tensor power spectrum dS}.} of $\Delta^2_h(k_\ast)\simeq3\times10^{-11}$ and a tensor-to-scalar ratio of $r(k_\ast)\simeq0.01$. 
This means that if the number of e-folds were $N_{\mathrm{obs}}\lesssim30$, we would get a tensor-to-scalar ratio inconsistent with current upper bounds, i.e. $r(k_\ast)\gtrsim0.07$ (this is shown as a gray shaded region in Figure \ref{fig: Hinf VS Nmin}). In other words, in this picture, dS needs to last long enough to amplify the scalar power spectrum with respect to the tensor one. Further insights can be found in Appendix \ref{sec: Decrease of the tensor-to-scalar ratio $r$}.
Moreover, for the reference value of $N_{\mathrm{obs}}=60$, i.e. for $k_{\rm end}\sim10^{22}\,\rm Mpc^{-1}$, we get $H_{\rm inf}\simeq2\times10^{10}\,\rm{GeV}$, $\Delta^2_h(k_\ast)\simeq10^{-17}$ and $r(k_\ast)\simeq5\times10^{-9}$. \\

Finally, we compute the dimensionless power spectrum of scalar perturbations $\Delta^2_\phi(k)$ as a function of the external wave-number $k$ in Figure \ref{fig: scalar and tensor PS VS Hinf N=30,60 together} (solid black line) and we compare it with the tensor power spectrum $\Delta^2_h(k)$. As a dashed blue line, we plot $\Delta^2_h(k)$ for the case $N_{\rm obs}=30$, while as a dotted red line, we show $\Delta^2_h(k)$ for the case $N_{\rm obs}=60$.  
From Figure \ref{fig: scalar power spectrum VS external wavenumber k Nmin=30,60}, we see that the power spectra are exactly scale-invariant. However, we argue that, if one takes into account that different modes spend different times outside the horizon, one can infer a red tilt for the scalar power spectrum, as we show in Appendix~\ref{sec:The scalar spectral tilt}.\\

\begin{figure}
    \centering  
    \includegraphics[width=0.45\textwidth]{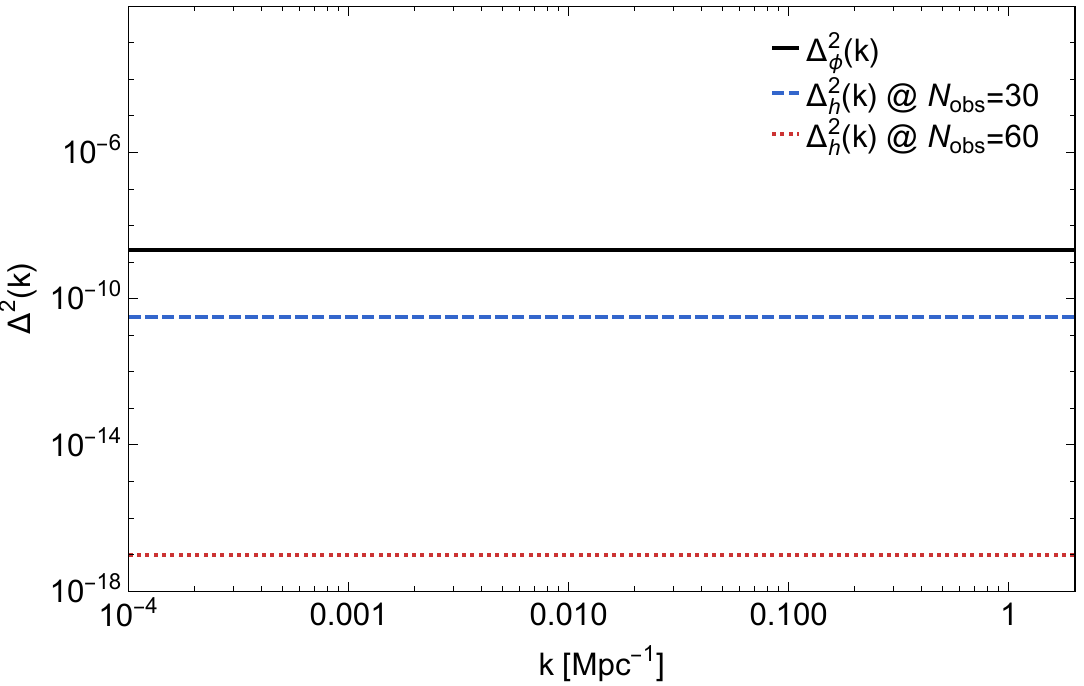} 
    \caption{Dimensionless scalar and tensor power spectra as a function of the external wavenumber $k$. $\Delta^2_\phi(k)$ is shown as a solid black line and it matches the observed level of scalar fluctuations at CMB scales \cite{Planck:2018jri,Planck:2018vyg}. As a dashed blue line, we plot $\Delta^2_h(k)$ for $N_{\rm obs}=30$. In this case, we fix the energy scale to $H_{\rm inf}\simeq~3\times~10^{13}\,\rm{GeV}$ to match \textit{Planck} constraints at $k_\ast$ \cite{Planck:2018jri,Planck:2018vyg}, as explained in the main text. Thus, the tensor-to-scalar ratio results $r(k_\ast)\simeq0.01$. The dotted red line shows $\Delta^2_h(k)$ for the case $N_{\rm obs}=60$, in which $H_{\rm inf}\simeq2\times10^{10}\,\rm{GeV}$ and $r(k_\ast)\simeq5\times10^{-9}$.}  
    \label{fig: scalar power spectrum VS external wavenumber k Nmin=30,60} 
\end{figure}

Before concluding this Section, it is worth making a general observation. The scale invariance observed in Figure~\ref{fig: scalar power spectrum VS external wavenumber k Nmin=30,60} may be interpreted as a non-linear effect connecting modes $k_1$ and $k_2$ with $k$. Such a correlation can arise, for instance, from a “scattering”-like mechanism that couples different modes or ranges of modes. A similar phenomenon has been discussed in Ref.~\cite{Dvali:2017eba}.

In general, for sufficiently intense gravitational fields or GWs such as those generated in extreme events like black hole mergers—the mutual interaction among GWs can effectively act as a mode coupling. Analogous behavior appears in condensed matter systems, for example, in superfluid condensates, where different phases interact through wave coupling. In such systems, the phenomenon can be modeled by considering how the constituent waves influence one another, particularly in regimes where multiple condensate phases coexist.\\
In our case, this type of coupling can occur within the horizon, either between different phases of a graviton condensate or through interactions between particles and gravitons. Formally, this amounts to including coupling terms in the system’s dynamical equations, an approach that bears a close analogy to Ginzburg–Landau theory, where phase interactions are described by coupled order parameters.

In our scenario, as the modes exit the horizon, they display this type of coupling. Consequently, the quantity $x_{\rm max} = k_{\rm end}/H_0 = e^{N_{\rm obs}}$ acquires a deeper physical meaning in the regime of strong gravity. Specifically, $x_{\rm max}$ represents the largest possible scale ratio, obtained by fixing the numerator to $k_{1{\rm max}} =k_{\rm end}= H_0 e^{N_{\rm obs}}$ and minimizing the denominator (i.e., setting $k = H_0$), which yields $x_{\rm max} = e^{N_{\rm obs}}$.\\

Note that considering instead $x_{\rm max}$ depending on $1/k$  would imply that the power spectrum decreases until it becomes red. This can be seen for the case where $k \to k_{\rm end} \sim 1/|\eta_{\rm end}|$. This can also be understood in the superfluid analogy,  where the condensed and phonon components are governed by coupled wave-kinetic equations (see, e.g., Ref.~\cite{Tsuzuki71}).

\subsection{End of de Sitter inflationary scenario}
\label{sec: End of de Sitter inflationary scenario}

At this point, it is interesting to connect the analysis we have done so far, in which we have treated the dS universe in a completely classical manner, with the approach adopted and studied by Ref. \cite{Dvali:2017eba}, in which a classical dS spacetime can be represented as the expectation value of a corresponding quantum graviton field across a well-defined coherent state. This perspective allows us to translate our approach into microscopic language, in which dS is treated as a quantum state, and ultimately, the existence of a finite quantum cut-off time provides a natural and harmonious existence for Guth's original inflationary scenario. In particular, an important consequence is the definition of the quantum break-time, which is the time scale after which the system can no longer be studied classically, and it is determined by \cite{Dvali:2017eba}
\begin{eqnarray}
    t_q=\dfrac{1}{\mathcal{N}_{\rm sp}}\dfrac{m_{\rm pl}^2}{H_{\rm inf}^3}\,,
    \label{eq: quantum break-time de sitter}
\end{eqnarray}
where $\mathcal{N}_{\rm sp}$ is the number of particle species during dS, and it is useful to connect it with the quantum break-time  $t_q$. 
In particular, as pointed out in Ref. \cite{Dvali:2017eba}, since the presence of more particle species opens up more channels for Gibbons-Hawking particle production, the quantum break-time of dS becomes $\mathcal{N}_{sp}$ times shorter (see Eq. \eqref{eq: quantum break-time de sitter}).
Following \cite{Dvali:2017eba}, note that $m_{\rm pl}^2/H_{\rm inf}^2$ is the critical number $\mathcal{N}_{\rm cr}$, i.e, the maximum number of allowed particle species that can yield a dS metric. Since we are assuming that the inflation period before $t_q$ can be treated classically, it cannot exceed the average graviton occupancy number of the coherent dS state. In our case, if $\mathcal{N}_{\rm sp}$ becomes larger than the critical number $\mathcal{N}_{\rm cr}=m_{\rm pl}^2/H_{\rm inf}^2$, it is inevitable that the dS quantum breaking time becomes smaller than the Hubble time $H_{\rm inf}^{-1}$. This leads to the behavior of our dS system in a non-classical system, with a consequent decay of inflation until reaching a regime of quantum gravity. Moreover, given $\mathcal{N}_{\rm sp}$, the longer the classical duration of inflation, the lower the energy scale of inflation, similarly to what we have seen in Figure \ref{fig: Hinf VS Nmin}. 
Since the Hubble parameter is constant in dS, the duration of inflation $t_{\rm end}$ can be written as $$t_  \mathrm{end}=\dfrac{N_\mathrm{tot}}{H_\mathrm{inf}}\,,$$ where the number of e-foldings is  $N_\mathrm{tot}\equiv\ln{(a_\mathrm{end}/a_\mathrm{in})}$.
Then, taking the quantum breaking time $t_q$ to coincide with the duration of inflation\footnote{Notice that here both $t_q$ and $t_{\rm end}$ are physical times.} $t_{\rm end}$, it is straightforward to show that the classical description of the inflationary epoch can only be trusted for a number of e-folds \cite{Dvali:2017eba}
\begin{eqnarray}
    N_\text{tot}<\dfrac{1}{\mathcal{N}_{sp}}\dfrac{m_{\rm pl}^2}{H_{\rm inf}^2}\,.
    \label{eq: upper bound efolds from Dvali}
\end{eqnarray}

\begin{figure}[t!]
    \centering  
    \includegraphics[width=0.48\textwidth]{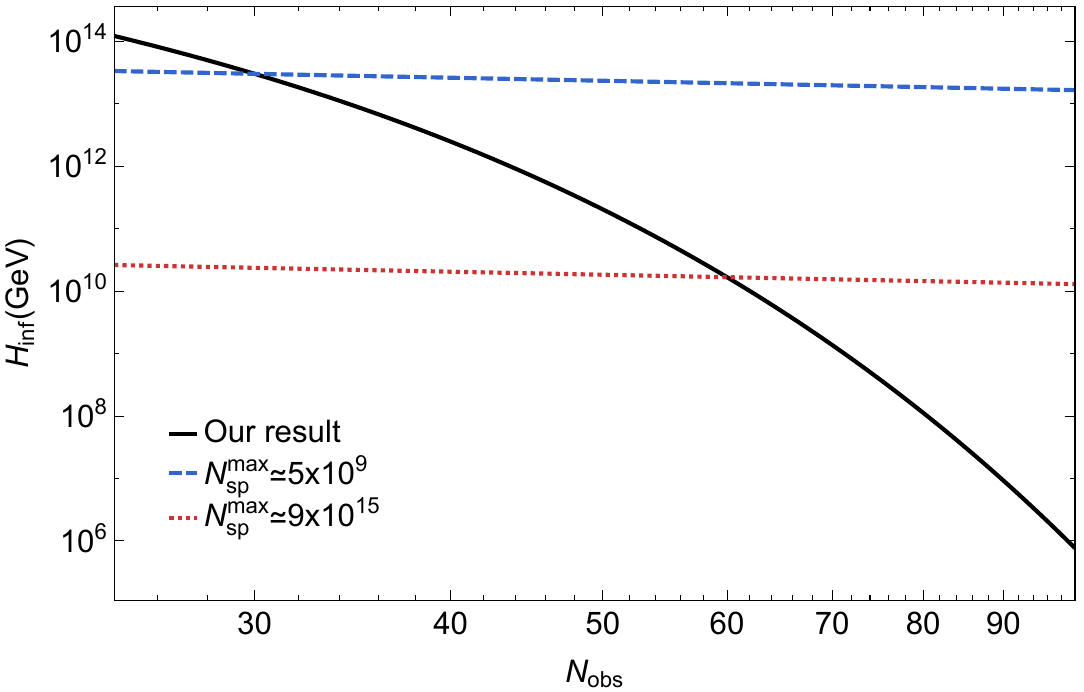}  
    \caption{Energy scale of inflation $H_{\rm inf}(\rm GeV)$ as a function of the minimum number of e-folds $N_{\rm obs}$. The solid black line represents our result from the numerical integration of Eq. \eqref{eq: dimensionless power spectrum to compute} (same as Figure \ref{fig: Hinf VS Nmin}). We plot the relation between $H_{\rm inf}$ and $N_{\rm obs}$ given by Eq. \eqref{eq: upper bound particle species} for different numbers of particle species $\mathcal{N}^{\rm max}_{\rm sp}\simeq5\times10^{9},9\times10^{15}$ with a dashed blue line and a dotted red line, respectively. By construction, the lines cross at $[N_{\rm obs},H_{\rm inf}(\mathrm{GeV})]\simeq[30, 3\times10^{13}\;\rm GeV]$ if $\mathcal{N}^{\rm max}_{\rm sp}\simeq~5\times10^{9}$, and at $[N_{\rm obs},H_{\rm inf}(\mathrm{GeV})]\simeq[60, 2\times10^{10}\;\rm GeV]$ if $\mathcal{N}^{\rm max}_{\rm sp}\simeq~9\times~10^{15}$.}  
    \label{fig: Hinf VS N our results and Dvali}  
\end{figure}

In our framework, we have seen that by fixing the observable number of e-folds\footnote{Remember that by fixing $N_{\rm obs}$, we fix the comoving scale leaving the horizon at the end of inflation, $k_{\rm end}$, which is related to the super-horizon requirement.} $N_{\mathrm{obs}}$ and matching the result for the scalar amplitude $\Delta_\phi^2(k_\ast)$ to the \textit{Planck} observational constraint at the CMB pivot scale, we can obtain $H_{\rm inf}$ and, consequently, $\Delta^2_h(k_\ast)$ and $r(k_\ast)$. We plot the energy scale of inflation $H_{\rm inf}$ as a function of $N_{\rm obs}$ in Figure \ref{fig: Hinf VS N our results and Dvali} as a solid black line (this is the same as Figure \ref{fig: Hinf VS Nmin} with the $x$-axis in log scale). $N_{\rm tot}$ in Eq. \eqref{eq: upper bound efolds from Dvali} and $N_{\rm obs}$ are simply related by the fact that $N_{\rm obs}\lesssim N_{\rm tot}$. 
If we take $N_{\rm obs}=N_{\rm tot}$, the quantum break-time leads to an upper bound on the number of light particle species $\mathcal{N}_{sp}$ allowed by the theory as 
\begin{eqnarray}
    \mathcal{N}_{\rm sp}<\dfrac{1}{N_{\rm obs}}\dfrac{m_{\rm pl}^2}{H_{\rm inf}^2}\equiv\mathcal{N}_{\rm sp}^{\rm max}\,,
    \label{eq: upper bound particle species}
\end{eqnarray}
where $\mathcal{N}_{\rm sp}^{\rm max}$ is the maximum number of species.
Using our results for $N_{\rm obs}$ and $H_{\rm inf}$, we can compute $\mathcal{N}_{\rm sp}^{\rm max}$. For instance, for $N_{\mathrm{obs}}=30$ and $H_{\rm inf}\simeq3\times10^{13}\;\rm GeV$, we obtain $\mathcal{N}_{\rm sp}\lesssim5\times10^{9}$; while $N_{\mathrm{obs}}=60$ and $H_{\rm inf}\simeq~2\times~10^{10}\;\rm GeV$ yield an upper bound on the number of species of $\mathcal{N}_{\rm sp}\lesssim9\times10^{15}$.
We can see this visually by plotting the energy scale of inflation as a function of $N_{\rm obs}$ for different number of particle species $\mathcal{N}_{\rm sp}^{\rm max}$ in Figure \ref{fig: Hinf VS N our results and Dvali}. More precisely, we plot the cases $\mathcal{N}_{\rm sp}^{\rm max}\simeq 5 \times 10^{9},9 \times 10^{15}$ with a dashed blue line and a dotted red line, respectively. 
From Figure \ref{fig: Hinf VS N our results and Dvali}, in agreement with Ref. \cite{Dvali:2017eba}, we see that the longer inflation lasts, the lower the bound on $H_{\rm inf}$ and the more species are allowed by the theory. Further remarks on the end of the dS scenario can be found in Appendix \ref{sec: End of de Sitter inflationary scenario appendix}.\\

Before concluding, it is worth making the following general observation. The dS quantum breaking time \eqref{eq: quantum break-time de sitter} exhibits a formal analogy with the evaporation time of black holes undergoing Hawking radiation \cite{1974Natur.248...30H,Page:1976df}. This correspondence, first emphasized in Ref. \cite{Dvali:2017eba}, reflects the thermodynamic resemblance between dS and black hole horizons, both radiating with a temperature proportional to their surface gravity—the Gibbons–Hawking temperature in the dS case \cite{PhysRevD.15.2738,PhysRevD.15.2752}. In this sense, the quantum break-time may be viewed as the characteristic lifetime of a Hubble patch before the dS horizon ``evaporates” \cite{Padmanabhan:2002ji,Dvali:2017eba,Markkanen:2017abw}. 
For further discussion on the thermodynamics of cosmological horizons in connection to the black hole case, see, e.g., Refs. \cite{Mottola:1984ar,Mottola:1985qt,Padmanabhan:2002ji,Padmanabhan:2002sha,Markkanen:2016jhg,Markkanen:2016vrp,Markkanen:2017abw,Alicki:2023rfv,Alicki:2023tfz}. 

\section{Conclusions}
\label{sec: Conclusions}

In this work, we investigated the predictions of the scenario introduced in Ref. \cite{Bertacca:2024zfb}, in which scalar perturbations are generated without invoking an inflaton field. In particular, we derived the power spectrum of scalar perturbations and assessed the phenomenological implications of this inflationary mechanism, namely, the duration and the energy scale of inflation, and the amplitude of the fluctuations. As in the case of our previous work~\cite{Bertacca:2024zfb}, the presence of a fluid term due to the graviton condensate allows the equations to be self-consistent. This, in turn, provides scalar perturbations that dominate over the tensor ones.

Our calculations provide the complete expression for the scalar power spectrum $\Delta^2_\phi(k)$ predicted by this scenario, and we have shown that it is possible to generate a scale-invariant spectrum. Notably, the model has only one free parameter: the energy scale of dS inflation, $H_{\mathrm{inf}}$. Given the number of e-folds $N_{\rm obs}$, this enables a direct connection between $H_{\mathrm{inf}}$ and the amplitude of the scalar fluctuations, which is tightly constrained at CMB scales \cite{Planck:2018jri,Planck:2018vyg}. 
Through numerical integration, we find that for $30-60$ e-folds, the energy scale $H_{\rm inf}\simeq5\times10^{13}\;\mathrm{GeV}-2\times10^{10}\;\mathrm{GeV}$ yields an amplitude of the scalar power spectrum in agreement with \textit{Planck} data, giving a tensor-to-scalar ratio $r\simeq0.01-5\times10^{-9}$.
Notice that the mechanism discussed in this paper to amplify the scalars with respect to the tensors complements the previous analysis in~\cite{Bertacca:2024zfb}. The complete study of this boosting mechanism will be dealt with in future work.

Moreover, we have shown how to connect our analysis to the concept of the quantum break-time of dS and the number of particle species. We have seen that the longer inflation lasts, the lower the bound on the energy scale and the more species it houses.\\
Furthermore, we argue that if we account for the time of horizon crossing—after which the scalar modes become constant—that depends on the comoving wavenumber $k$, we can obtain a red-tilted spectrum. This point will be studied in more detail in a forthcoming paper.\\
As already pointed out in \cite{Bertacca:2024zfb}, this inflationary mechanism offers a promising model-independent picture of inflation. A distinguishing feature would be its intrinsic non-Gaussianity, which we intend to explore in a future publication.

As can be seen from this study, if the energy scale and the total number of species in de Sitter are determined, i.e., the intrinsic characteristics of de Sitter, then this would explain why the amplitude of the scalar fluctuations has the value it has. This finding will have far-reaching consequences as it would explain one of the fundamental values of the current cosmological model, which is currently unexplained.

\begin{acknowledgments}
The authors acknowledge the useful discussions and comments received from P. Tejerina and L. Sarieddine.
The project that gave rise to these results received the support of a fellowship from “la Caixa” Foundation (ID 100010434), awarded to MT with code LCF/BQ/DI24/12070012.
The work of RJ is supported by the Simons Foundation.
Funding for the work of MT and RJ was partially provided by project PID2022-141125NB-I00, and grant CEX2024-001451-M funded by MICI-U/AEI/10.13039/501100011033. 
DB and SM acknowledge support from the COSMOS network (www.cosmosnet.it) through ASI (Italian Space Agency) Grants 2016-24-H.0, 2016-24-H.1-2018 and 2020-9-HH.0.
\end{acknowledgments}

\appendix
\section{Fourier space conventions}
\label{sec:Fourier space conventions}

In this Appendix, we present our Fourier conventions and the definition of the polarization tensors.

The Fourier transform of a scalar quantity, $\phi(\mathbf{x})$, can be expressed as 
\begin{eqnarray}
    \phi(\mathbf{x})=\frac{1}{(2\pi)^{3}} \int d^3\mathbf{k}\, \phi(\mathbf{k})\,e^{i\mathbf{k}\cdot \mathbf{x}}\;.
\end{eqnarray}
Similarly, we decompose the symmetric transverse-traceless tensor $\chi_{ij}(\mathbf{x})$ into plane waves with different magnitudes and orientations such as
\begin{eqnarray}
\chi_{ij}(\mathbf{x}) =\frac{1}{(2\pi)^3} \int d^3\mathbf{k}\, \sum_\lambda  \chi_\lambda(\mathbf{k}) \,e^\lambda_{ij}(\mathbf{k})\, e^{i\mathbf{k}\cdot \mathbf{x}} \;,
\end{eqnarray}
where $\lambda$ is the polarization index and $e^\lambda_{ij}(\mathbf{k})$ are the time-independent polarization tensors.\\
For our purposes, it is convenient to use the so-called circular left (L) and right (R) polarization states, which are constructed from a linear combination of the plus (+) and cross ($\times$) polarizations as:
\begin{eqnarray}
        e^R_{ij}(\mathbf{k}) &= \frac{1}{\sqrt{2}} \left ( e^+_{ij}(\mathbf{k}) + i e^{\times}_{ij}(\mathbf{k})  \right ) \;,  \\
        e^L_{ij}(\mathbf{k}) &= \frac{1}{\sqrt{2}} \left ( e^+_{ij}(\mathbf{k}) - i e^{\times}_{ij}(\mathbf{k})  \right )\;.
\end{eqnarray}
It can be shown that the following relations hold
\begin{eqnarray}
        e^L_{ij}(\mathbf{k})e_L^{ij}(\mathbf{k})=e^R_{ij}(\mathbf{k})e_R^{ij}(\mathbf{k})=0\;,\nonumber\\
        e_{ij}^{L}(\mathbf{k}) e^{ij}_R(\mathbf{k}) = 2 \;, \nonumber\\
        e_{ij}^{\ast\lambda}(\mathbf{k})=e_{ij}^{\lambda}(-\mathbf{k})= e_{ij}^{-\lambda}(\mathbf{k}) \;,\nonumber\\
        \chi^\ast_\lambda(\mathbf{k})=\chi_\lambda(-\mathbf{k})\;,
\end{eqnarray}
where $\lambda=L,R$ and $-\lambda$ means simply $-L=R$ and $-R=L$. \\
In the main text, we work with three momenta, i.e., $\{\mathbf{k}_1,\mathbf{k}_2,\mathbf{k}\}$, that form a triangle in momentum space because of $\delta^{(3)}(\mathbf{k}-(\mathbf{k}_1+\mathbf{k}_2))$.
Due to isotropy, we are free to place the triangle in the $(x,y)$-plane without loss of generality. To do so, we align $\mathbf{k}_1$ with the $x$-axis, therefore we have 
\begin{eqnarray}
\label{def k_1}
    \mathbf{k}_1=k_1(1,0,0)\;.
\end{eqnarray} 
Then, the vectors $\mathbf{k}_2$ and $\mathbf{k}$ read
\begin{eqnarray}
\label{def k2}
    \mathbf{k}_2 = k_2 \left (\cos{\theta}, \sin{\theta}, 0 \right )\;,
\end{eqnarray}
\begin{eqnarray}
\label{def k}
    \mathbf{k}=k(\cos{\Phi},\sin{\Phi},0)\;,
\end{eqnarray}
where $\theta$ and $\Phi$ are the angles that $\mathbf{k}_2$ and $\mathbf{k}$ form with $\mathbf{k}_1$, respectively.
Hence, the scalar and vector products between $\mathbf{k}_1$ and $\mathbf{k}_2$ are simply
\begin{eqnarray}
\label{scalar product k1 k2}
\mathbf{k}_1\cdot\mathbf{k}_2=k_1k_2\cos{\theta}\;,
\end{eqnarray}
and
\begin{eqnarray}
    \label{cross product k1 k2}
    \mathbf{k}_1\times\mathbf{k}_2=(0,0,k_1k_2\sin{\theta})\;.
\end{eqnarray}

From the cosine theorem, we get the expression for $\cos{\theta}$ as a function of the wavenumbers:
\begin{eqnarray}
    \cos{\theta}=\dfrac{k_1^2+k_2^2-k^2}{2k_1k_2}\;.
    \label{eq: cos(theta) cosine theorem}
\end{eqnarray}
Moreover, with these choices of the momenta, we can obtain the explicit expressions for the L and R polarization tensors (see, e.g., Refs. \cite{Soda:2011am,Bartolo:2017szm}).
We have that
\begin{eqnarray}
\mathbf{e}^\lambda(\mathbf{k}_1)=\dfrac{1}{\sqrt{2}}
    \begin{pmatrix}
0 & 0 & 0\\
0 & 1 & is_\lambda\\
0 & is_\lambda & -1
\end{pmatrix}\;,
\label{polarization tensor k1}
\end{eqnarray}
where $s_L=-1$ and $s_R=+1$. To obtain $\mathbf{e}^\lambda(\mathbf{k}_2)$ we just need to rotate \eqref{polarization tensor k1} by an angle $\theta$ as
\begin{eqnarray}
\mathbf{e}^\lambda(\mathbf{k}_2)=\dfrac{1}{\sqrt{2}}
    \begin{pmatrix}
\sin^2{\theta} & -\sin{\theta}\cos{\theta} & -is_\lambda\sin{\theta}\\
-\sin{\theta}\cos{\theta} & \cos^2{\theta} & is_\lambda\cos{\theta}\\
-is_\lambda\sin{\theta} & is_\lambda\cos{\theta} & -1
\end{pmatrix}\,,\quad\quad\quad
\label{polarization tensor k2}
\end{eqnarray}
and similarly for $\mathbf{e}^\lambda(\mathbf{k})$.\\
Finally, through an explicit calculation, we can show that the following expressions hold:
\begin{eqnarray}
\label{relation eLeL and eReR}
    e^L_{ij}(\mathbf{k}_1)e^{ij}_L(\mathbf{k}_2)= e^R_{ij}(\mathbf{k}_1)e^{ij}_R(\mathbf{k}_2)=\dfrac{1}{2}(1-\cos{\theta})^2\;,\quad\quad\quad
\end{eqnarray}
\begin{eqnarray}
\label{relation eLeR and eReL}
    e^L_{ij}(\mathbf{k}_1)e^{ij}_R(\mathbf{k}_2)=e^R_{ij}(\mathbf{k}_1)e^{ij}_L(\mathbf{k}_2)=\dfrac{1}{2}(1+\cos{\theta})^2\;,\quad\quad\quad
\end{eqnarray}
and
\begin{eqnarray}
\label{relation k1 e1 k2 e2}
     k_{1\,i}\, e^{lj}_{\lambda_1}(\mathbf{k_1}) k_{2\,l} e^i_{\lambda_2\,j}(\mathbf{k}_2)=\dfrac{1}{2}k_1k_2\sin^2{\theta}(s_{\lambda_1}s_{\lambda_2}-\cos{\theta})\;.\quad\quad\quad
\end{eqnarray}
\\
\section{Power Spectrum computation}
\label{sec:Power Spectrum computation}
We write in Fourier space the particular solution \\$\phi_2=\mathcal{F}_\chi/4$, with $\mathcal{F}_\chi$ given by Eq. \eqref{eq:F_chi}.
For instance, if we consider only the first terms in \eqref{eq:F_chi}, i.e., the ones collectively called as 'A', we obtain
\begin{eqnarray}
\label{phi_2 fourier space first part}
    \phi^{(A)}_2(\mathbf{k})&=&-\dfrac{1}{k^2}\sum_{\lambda_1,\lambda_2}\int \dfrac{d^3\mathbf{k_1}}{(2\pi)^3}\int \dfrac{d^3\mathbf{k_2}}{(2\pi)^3}\delta^{(3)}(\mathbf{k}-(\mathbf{k}_1+\mathbf{k}_2))\nonumber\\
   &&\times Q^{(A)}_{\lambda_1,\lambda_2}(\mathbf{k}_1,\mathbf{k}_2)\,\chi_{\lambda_1}(\mathbf{k}_1)\chi_{\lambda_2}(\mathbf{k}_2)\;,\quad\quad\quad
    \end{eqnarray}
where we are summing over the two polarizations $\lambda_1,\lambda_2 = R, L $ and we have defined the function $Q^{(A)}_{\lambda_1,\lambda_2}(\mathbf{k}_1,\mathbf{k}_2)$ encoding the polarization tensors as
\begin{eqnarray}
    Q^{(A)}_{\lambda_1,\lambda_2}(\mathbf{k}_1,\mathbf{k}_2)&=&\dfrac{1}{4}(-k^2_1-k^2_2-3\mathbf{k}_1\cdot\mathbf{k}_2)\left[e^{ij}_{\lambda_1}(\mathbf{k_1})e_{ij}^{\lambda_2}(\mathbf{k}_2)\right]\nonumber\\
    &&+\dfrac{1}{2}k_{1\,i}\, e^{lj}_{\lambda_1}(\mathbf{k_1}) k_{2\,l} e^i_{\lambda_2\,j}(\mathbf{k}_2)\;.
    \end{eqnarray}
In the above expression, $e^{ij}_{\lambda_1}(\mathbf{k_1})e_{ij}^{\lambda_2}(\mathbf{k}_2)$ can be written explicitly using Eqs. \eqref{relation eLeL and eReR} and \eqref{relation eLeR and eReL}, while $k_{1\,i}\, e^{lj}_{\lambda_1}(\mathbf{k_1}) k_{2\,l} e^i_{\lambda_2\,j}(\mathbf{k}_2)$ is given by Eq. \eqref{relation k1 e1 k2 e2}.\\
To obtain the power spectrum $\mathcal{P}^{(AA)}_\phi(k)$, we substitute Eq. \eqref{phi_2 fourier space first part} into the two-point function \eqref{power spectrum definition}
\begin{widetext}
    \begin{eqnarray}
  \langle \phi^{(A)}(\mathbf{k}) \phi^{(A)}(\mathbf{k'}) \rangle &=&      \dfrac{1}{4}\langle\phi^{(A)}_2(\mathbf{k})\phi^{(A)}_2(\mathbf{k}')\rangle=\nonumber\\
  &=&\dfrac{1}{4}\sum_{\lambda_1,\lambda_2}\dfrac{1}{k^2k^{'2}}\int \dfrac{d^3\mathbf{k_1}}{(2\pi)^3}\int \dfrac{d^3\mathbf{k_2}}{(2\pi)^3}\int \dfrac{d^3\mathbf{k'_1}}{(2\pi)^3}\int \dfrac{d^3\mathbf{k'_2}}{(2\pi)^3}
    \delta^{(3)}(\mathbf{k}-(\mathbf{k}_1+\mathbf{k}_2))\delta^{(3)}(\mathbf{k}'-(\mathbf{k}'_1+\mathbf{k}'_2))\nonumber\\
&&\times Q^{(A)}_{\lambda_1,\lambda_2}(\mathbf{k}_1,\mathbf{k}_2)Q^{(A)}_{\lambda'_1,\lambda'_2}(\mathbf{k}'_1,\mathbf{k}'_2)\langle\chi_{\lambda_1}(\mathbf{k}_1)\chi_{\lambda_2}(\mathbf{k}_2)\chi_{\lambda'_1}(\mathbf{k}'_1)\chi_{\lambda'_2}(\mathbf{k}'_2)\rangle\;.
    \label{calculation first part PS}
\end{eqnarray}
\end{widetext}
In the expression above, we need to evaluate the four-point function of the first-order tensor perturbations~$\chi_{\lambda_i}(\mathbf{k}_i)$.
At leading order, this can be written in terms of products of two-point correlations\\(Wick's theorem):
\begin{eqnarray}
\label{four point function}
    &\langle \chi_{\lambda_1}(\mathbf{k}_1)\chi_{\lambda_2}(\mathbf{k}_2)\chi_{\lambda'_1}(\mathbf{k}'_1)\chi_{\lambda'_2}(\mathbf{k}'_2)\rangle = \nonumber\\
    &=\langle \chi_{\lambda_1}(\mathbf{k}_1)\chi_{\lambda_2}(\mathbf{k}_2)\rangle \langle\chi_{\lambda'_1}(\mathbf{k}'_1)\chi_{\lambda'_2}(\mathbf{k}'_2)\rangle \nonumber \\
    &+ \langle \chi_{\lambda_1}(\mathbf{k}_1)\chi_{\lambda'_1}(\mathbf{k}'_1)\rangle \langle\chi_{\lambda_2}(\mathbf{k}_2)\chi_{\lambda'_2}(\mathbf{k}'_2)\rangle \nonumber\\
    &+\langle \chi_{\lambda_1}(\mathbf{k}_1)\chi_{\lambda'_2}(\mathbf{k}'_2)\rangle \langle\chi_{\lambda_2}(\mathbf{k}_2)\chi_{\lambda'_1}(\mathbf{k}'_1)\rangle \;, 
\end{eqnarray}
where the first part is vanishing because they are disconnected. The two-point correlations are given by
\begin{eqnarray}
    \langle \chi_{\lambda_1}(\mathbf{k}_1)\chi_{\lambda_2}(\mathbf{k}_2)\rangle=(2\pi)^3\delta_{\lambda_1\lambda_2}\delta^{(3)}(\mathbf{k}_1+\mathbf{k}_2)\mathcal{P}_\chi(k_1)\;,\quad\quad\quad
\end{eqnarray}
where $\mathcal{P}_\chi(k)={1\over2}\mathcal{P}_h(k)$ and $\mathcal{P}_h(k)$ the tensor power spectrum given in Eq. \eqref{tensor power spectrum dS}. Moreover, since the integrand in Eq. \eqref{calculation first part PS} is invariant under the interchange of $\mathbf{k}_1\leftrightarrow\mathbf{k}_2=\mathbf{k}-\mathbf{k}_1$ (see e.g. Ref. \cite{Espinosa:2018eve}), the two non-trivial terms in \eqref{four point function} yield the same contribution, thus we can compute one of them and multiply the result by 2.\\
When we evaluate explicitly \eqref{calculation first part PS} using the properties of the polarization tensors, rewriting the cosine and sine terms using Eqs. \eqref{scalar product k1 k2} and \eqref{cross product k1 k2}, and summing over the two polarizations, we obtain the expression \eqref{PS first part- first part} for the primordial power spectrum. 

Now, let us consider the computation of the Fourier transform of the 'B' term in Eq. \eqref{eq:F_chi}, i.e. $-6\nabla^{-4}\partial_i\partial^j\mathcal{A}^i_j$, with $\mathcal{A}^i_j$ defined in \eqref{eq:calA}. This gives
\begin{eqnarray}
\phi^{(B)}_2(\mathbf{k})&=&\dfrac{1}{k^4}\sum_{\lambda_1,\lambda_2}\int \dfrac{d^3\mathbf{k_1}}{(2\pi)^3}\int \dfrac{d^3\mathbf{k_2}}{(2\pi)^3}\delta^{(3)}(\mathbf{k}-(\mathbf{k}_1+\mathbf{k}_2))\nonumber\\
&&\times Q^{(B)}_{\lambda_1\lambda_2}(\mathbf{k}_1,\mathbf{k}_2)\chi_{\lambda_1}(\mathbf{k}_1)\chi_{\lambda_2}(\mathbf{k}_2)\;,
    \label{phi_2 fourier space second part}
\end{eqnarray}
where we have defined the function $Q^{(B)}_{\lambda_1\lambda_2}(\mathbf{k}_1,\mathbf{k}_2)$ encoding the polarization tensors as
\begin{widetext}
    \begin{eqnarray}
       Q^{(B)}_{\lambda_1\lambda_2}(\mathbf{k}_1,\mathbf{k}_2)&=& -\dfrac{3}{2}\left\{\dfrac{1}{2}\left[k_1^2(\mathbf{k}_1\cdot\mathbf{k}_2)+(\mathbf{k}_1\cdot\mathbf{k}_2)^2+k_1^2k_2^2+k_2^2(\mathbf{k}_1\cdot\mathbf{k}_2)\right]\epsilon^{\lambda_1\,lm}(\mathbf{k}_1)\epsilon^{\lambda_2}_{ml}(\mathbf{k}_2)\right.\nonumber\\
&&\left.+\left[(\mathbf{k}_1\cdot\mathbf{k}_2)^2+(k_1^2+k_2^2)(\mathbf{k}_1\cdot\mathbf{k}_2)+\dfrac{1}{2}k_1^4+\dfrac{1}{2}k_2^4\right]\epsilon^{\lambda_1\,ml}(\mathbf{k}_1)\epsilon^{\lambda_2}_{lm}(\mathbf{k}_2)\right.\nonumber\\
&&\left.-\left[(\mathbf{k}_1\cdot\mathbf{k}_2)+\dfrac{1}{2}k_1^2+\dfrac{1}{2}k_2^2\right]k_{1\,i}k_{2\,m}\epsilon^{\lambda_1\,ml}(\mathbf{k}_1)\epsilon^{\lambda_2\,i}_{\,l}(\mathbf{k}_2)-\left[(\mathbf{k}_1\cdot\mathbf{k}_2)+\dfrac{1}{2}k_1^2+\dfrac{1}{2}k_2^2\right]k_1^jk_{2\,m}\epsilon^{\lambda_1\,ml}(\mathbf{k}_1)\epsilon^{\lambda_2}_{lj}(\mathbf{k}_2)\right.\nonumber\\
&&\left.+k_{1\,i}k_1^jk_{2\,m}k_{2\,l}\epsilon^{\lambda_1\,ml}(\mathbf{k}_1)\epsilon^{\lambda_2\,i}_{\,j}(\mathbf{k}_2)+(\mathbf{k}_1\cdot\mathbf{k}_2)k_1^jk_{2\,i}\epsilon^{\lambda_1\,im}(\mathbf{k}_1)\epsilon^{\lambda_2}_{jm}(\mathbf{k}_2)-k_{1\,l}k_1^jk_{2\,m}k_{2\,i}\epsilon^{s\,mi}(\mathbf{k}_1)\epsilon^{\lambda_2\,l}_{\,j}(\mathbf{k}_2)\right\}\,.\quad\quad\;
    \end{eqnarray}
\end{widetext}
Then, the power spectrum $\mathcal{P}^{(BB)}(k)$, $\mathcal{P}^{(AB)}(k)$ and $\mathcal{P}^{(BA)}(k)$ can be extracted in a similar way as done for $\mathcal{P}^{(AA)}(k)$ and we get the expressions \eqref{PS second part- second part} and \eqref{PS first part- second part} reported in the main text.

\section{Full kernel}
\label{sec: Full kernel}
The kernels $\mathcal{K}^{(AA)}_h(\mathbf{k}_1,\mathbf{k}_2,k^2)$, $\mathcal{K}^{(BB)}_h(\mathbf{k}_1,\mathbf{k}_2,k^2)$ and $\mathcal{K}^{(AB)}_h(\mathbf{k}_1,\mathbf{k}_2,k^2)=\mathcal{K}^{(BA)}_h(\mathbf{k}_1,\mathbf{k}_2,k^2)$ are given in Eqs. \eqref{eq: kernel first part - first part}, \eqref{eq: kernel second part - second part}, and \eqref{eq: kernel first part - second part}, respectively, as a function of the wave-numbers in the main text.
For completeness, we provide here the expressions for the kernels in the $\{x,y\}$ variables introduced in Eq. \eqref{eq: def dimensionless variables x and y}. Together with the correct numerical factors, they read

\begin{widetext}
    \begin{eqnarray}
    \mathcal{K}^{(AA)}_h(x,y)&=\dfrac{1}{64\pi^2x^4y^4}&\left[x^{12}+(-1+y^2)^4(1+y^2)^2-2x^{10}(1+9y^2)+x^8(-1+54y^2+63y^4)\right.\nonumber\\
    &&\left.-2x^2(-1+y^2)^2(1-49y^2-9y^4+9y^6)+x^6(4+44y^2-564y^4+420y^6)\right.\nonumber\\
    &&\left.+x^4(-1-180y^2+746y^4-564y^6+63y^8)\right]\;,
    \label{eq: kernel AA}
    \end{eqnarray}

\begin{eqnarray}
    \mathcal{K}^{(BB)}_h(x,y)&=\dfrac{9}{256\pi^2x^4y^4}&\left[9x^{16}+4x^{14}(-9+22y^2)+(-1+y^2)^4(1+3y^4)^2-4x^{12}(-15+59y^2+49y^4)\right.\nonumber\\
    &&\left.-4x^{10}(15-122y^2+25y^4+38y^6)+x^8(46-652y^2+580y^4+372y^6+502y^8)\right.\nonumber\\
    &&\left.+4x^2(-1+y^2)^2(-1+28y^2-8y^4+70y^6-15y^8+22y^{10})\right.\nonumber\\
    &&\left.-4x^6(7-144y^2+206y^4+52y^6-93y^8+38y^{10})\right.\nonumber\\
    &&\left.-4x^4(-3+65y^2-261y^4+206y^6-145y^8+25y^{10}+49y^{12})\right]\;,
    \label{eq: kernel BB}
\end{eqnarray}
\begin{eqnarray}
    \mathcal{K}^{(AB)}_h(x,y)=\mathcal{K}^{(BA)}_h(x,y)&=&-\dfrac{3}{128\pi^2x^4y^4}\left[3x^{14}-x^{12}(9+23y^2)+x^{10}(7+158y^2-205y^4)\right.\nonumber\\
    &&\left.+(-1+y^2)^4(1+y^2)(1+3y^4)-x^2(-1+y^2)^2(3-104y^2-2y^4-122y^6+23y^8)\right.\nonumber\\
    &&\left.+x^6(-7+212y^2-658y^4+36y^6+225y^8)+x^8(3+5y^2(-49+69y^2+45y^4))\right.\nonumber\\
    &&\left.+x^4(5-209y^2+850y^4-658y^6+345y^8-205y^{10})\right]\;.
    \label{eq: kernel AB e BA}
\end{eqnarray}
Hence, the full kernel appearing in Eq. \eqref{eq: dimensionless power spectrum to compute} can be computed as
\begin{eqnarray}
     \mathcal{K}_h(x,y)= \mathcal{K}^{(AA)}_h(x,y)+ \mathcal{K}^{(BB)}_h(x,y)+2\mathcal{K}^{(AB)}_h(x,y)\;.
     \label{eq: def full kernel}
\end{eqnarray}
\end{widetext}

\section{Further Insights}
\label{sec: Further Insights}

\subsection{Decrease of the tensor-to-scalar ratio $r$}
\label{sec: Decrease of the tensor-to-scalar ratio $r$}
In this Section, we illustrate in another way the fact that we need inflation to last long enough in order to amplify scalars with respect to tensors.
To do so, given a value of $N_{\rm obs}$, we numerically integrate Eq. \eqref{eq: dimensionless power spectrum to compute} at the pivot scale $k_\ast=0.01\;\rm Mpc^{-1}$. Then, instead of fixing $H_{\rm inf}$ to match the amplitude of scalar perturbation constrained by \textit{Planck} \cite{Planck:2018jri,Planck:2018vyg}, we compute the scalar and tensor power spectra varying the energy scale $H_{\rm inf}$. 
We plot the results for $N_{\rm obs}=30,60$ in Figure \ref{fig: scalar and tensor PS VS Hinf N=30,60 together}. The solid red line represents the tensor power spectrum $\Delta^2_h(k_\ast)$ in dS space-time, which is defined in Eq. \eqref{tensor power spectrum dS}.
The dashed blue line is the scalar power spectrum for $N_{\rm obs}=30$, while the dotted blue line is $\Delta^2_\phi(k_\ast)$ for $N_{\rm obs}=60$. 
The two blue points show the scalar amplitude constrained by \textit{Planck} \cite{Planck:2018jri,Planck:2018vyg}, i.e. $\Delta^2_\phi(k_{\rm CMB}=k_\ast)\simeq2.1\times 10^{-9}$.
For the case $N_{\rm obs}=30$, $\Delta^2_\phi(k_{\rm CMB})$ corresponds to  $H_{\rm inf}\simeq3\times10^{13}\;\rm GeV$, thereby giving a tensor amplitude and a tensor-to-scalar ratio of $\Delta^2_h(k_\ast)\simeq3\times10^{-11}$ and $r(k_\ast)\simeq0.01$, respectively. 
For the case $N_{\rm obs}=60$, $\Delta^2_\phi(k_{\rm CMB})$ corresponds to  $H_{\rm inf}\simeq2\times10^{10}\;\rm GeV$, therefore we have that $\Delta^2_h(k_\ast)\simeq 10^{-17}$, and $r(k_\ast)\simeq5\times10^{-9}$.
\begin{figure}[htp!]
    \centering  
    \includegraphics[width=0.49\textwidth]{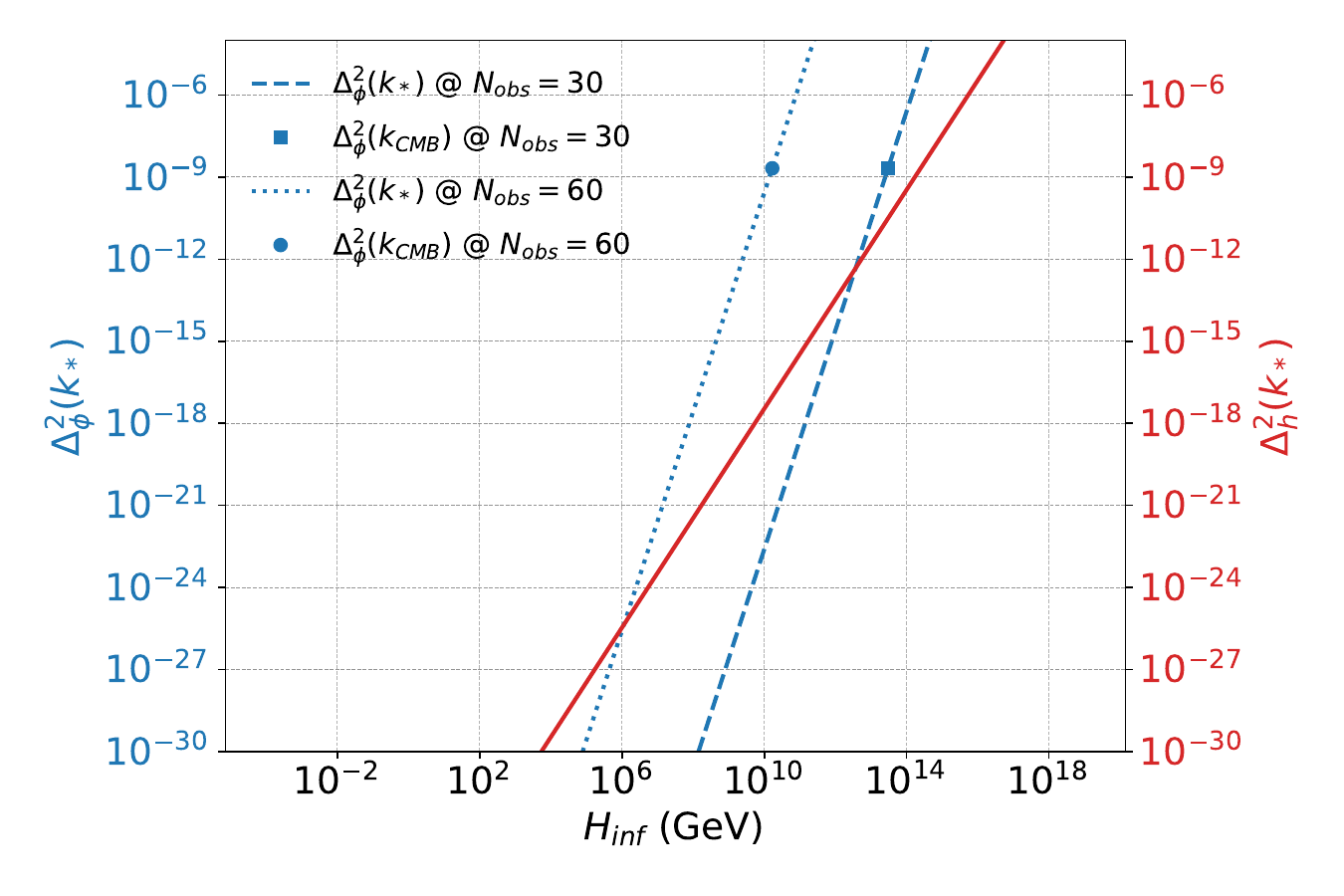}  
    \caption{Dimensionless scalar and tensor power spectra as a function of the energy scale $H_{\rm inf}$. 
    We fix $N_{\rm obs}$ and we obtain the scalar power spectrum $\Delta_\phi^2(k_\ast)$ from Eq. \eqref{eq: dimensionless power spectrum to compute} for different values of $H_{\rm inf}$. 
    As a dashed blue line, we show $\Delta_\phi^2(k_\ast)$ for $N_{\rm obs}=30$, as a dotted blue line, we plot $\Delta_\phi^2(k_\ast)$ for $N_{\rm obs}=60$, while the solid red line is the result for the tensor power spectrum $\Delta_h^2(k_\ast)$. We set the pivot scale to $k_\ast=~0.01\;\rm Mpc^{-1}$.
    The blue points correspond to the scalar amplitude $\Delta^2_\phi(k_{\rm CMB}=k_\ast)\simeq2.1\times10^{-9}$ constrained by \textit{Planck} \cite{Planck:2018jri,Planck:2018vyg}.
    For the case $N_{\rm obs}=30$, $\Delta^2_\phi(k_{\rm CMB})$ corresponds to  $H_{\rm inf}\simeq3\times10^{13}\;\rm GeV$, therefore the tensor power spectrum and the tensor-to-scalar ratio are $\Delta^2_h(k_\ast)\simeq~3\times~10^{-11}$ and $r(k_\ast)\simeq0.01$, respectively. 
    For the case $N_{\rm obs}=60$, $\Delta^2_\phi(k_{\rm CMB})$ corresponds to $H_{\rm inf}\simeq2\times10^{10}\;\rm GeV$, therefore $\Delta^2_h(k_\ast)\simeq 10^{-17}$, and $r(k_\ast)\simeq5\times10^{-9}$.
    }  
    \label{fig: scalar and tensor PS VS Hinf N=30,60 together} 
\end{figure}
We notice that if the energy scale of inflation is low enough, we obtain a scalar power spectrum  $\Delta_\phi^2\sim\Delta_h^2\Delta_h^2$, but, as the energy scale grows, we reach a turning point at which the two power spectra are the same, so $r=1$. Therefore, depending on the value of $H_{\rm inf}$, we obtain different values of the tensor-to-scalar ratio $r$, going from $r>1$ at the left side of the turning point to $r<1$.
For instance, notice that for $N_{\rm obs}=30$, the blue point corresponding to the scalar amplitude constrained by observations lies on the right side of the turning point, where $r<1$.
However, if we do the same procedure for $N_{\rm obs}\lesssim30$, we would obtain a tensor-to-scalar ratio inconsistent with current upper bounds, meaning that the energy scale matching the CMB constraints for $\Delta^2_\phi(k_\ast)$ would lie in the region where $r>1$.
On the other hand, if inflation lasts more e-folds, the results for $N_{\rm obs}=60$ show that the turning point shifts to lower values of $H_{\rm inf}$. Therefore, we need a lower energy scale to obtain the observed scalar amplitude, and we are in the region where $r\ll1$.

\subsection{End of de Sitter inflationary scenario}
\label{sec: End of de Sitter inflationary scenario appendix}

\begin{figure}
    \centering  
    \includegraphics[width=0.48\textwidth]{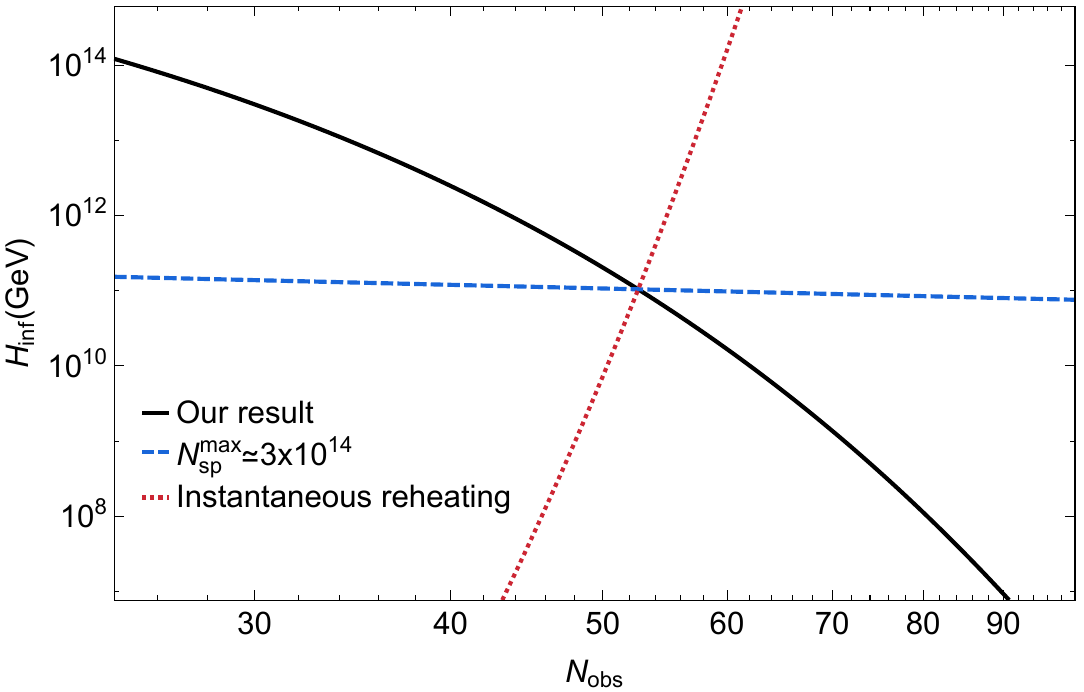}  
    \caption{
    Energy scale of inflation $H_{\rm inf}(\rm GeV)$ as a function of the minimum number of e-folds $N_{\rm obs}$. The solid black line represents our result from the numerical integration of Eq. \eqref{eq: dimensionless power spectrum to compute}. We plot as a dotted red line the number of e-foldings for the case of instantaneous reheating, i.e., Eq. \eqref{eq: efolds instantaneous reheating}. This crosses our result at $[N_{\rm obs},H_{\rm inf}(\mathrm{GeV})]\simeq[53, 10^{11}\;\rm GeV]$ and corresponds to a maximum number of particle species of $\mathcal{N}^{\rm max}_{\rm sp}\simeq3\times10^{14}$ using Eq. \eqref{eq: upper bound particle species} (dashed blue line).}
    \label{fig: Hinf VS N our results and Dvali reheating}  
\end{figure}
\begin{figure}
    \centering  
    \includegraphics[width=0.48\textwidth]{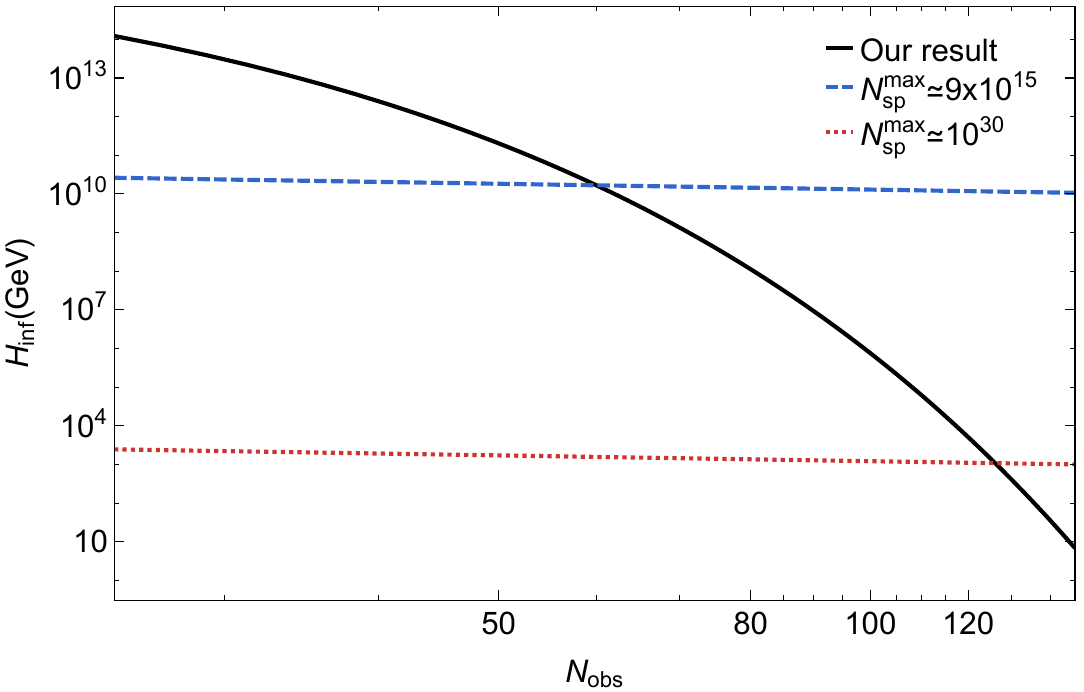}  
\caption{Energy scale of inflation $H_{\rm inf}(\rm GeV)$ as a function of the minimum number of e-folds $N_{\rm obs}$. The solid black line represents our result from the numerical integration of Eq. \eqref{eq: dimensionless power spectrum to compute}. We plot the relation between $H_{\rm inf}$ and $N_{\rm obs}$ given by Eq. \eqref{eq: upper bound particle species} for different numbers of particle species.
The dashed blue line shows the case $[N_{\rm obs},H_{\rm inf}(\mathrm{GeV})]\simeq[60, 2\times10^{10}\;\rm GeV]$ if $\mathcal{N}^{\rm max}_{\rm sp}\simeq9\times10^{15}$ discussed in the main text.
As a dotted red line, we plot the case $\mathcal{N}^{\rm max}_{\rm sp}\simeq10^{30}$, which intersects our result at $[N_{\rm obs},H_{\rm inf}(\mathrm{GeV})]\simeq[126,10^{3}\;\rm GeV]$.}  
\label{fig: Hinf VS N our results and Dvali Nsp=1e30}  
\end{figure}

In this Section, we give further details related to the end of dS inflationary scenario discussed in Sec.~\ref{sec: End of de Sitter inflationary scenario}.

First, we want to confront our results for $H_{\rm inf}$ and $N_{\rm obs}$ with the case of instantaneous reheating.
In this case, the number of e-foldings is \cite{Planck:2018jri} 
\begin{eqnarray}
    N_k=34.2+\ln{\dfrac{H_{\rm inf}}{1\,\mathrm{TeV}}}\;.
\label{eq: efolds instantaneous reheating}
\end{eqnarray}
Inverting the expression above, we plot the energy scale $H_{\rm inf}$ as a function of $N_k$ as a dotted red line in Figure \ref{fig: Hinf VS N our results and Dvali reheating}.
The line of instantaneous reheating intersects our result from Eq. \eqref{eq: dimensionless power spectrum to compute} (solid black line) at $N_{\rm obs}\simeq53$, which corresponds to $H_{\rm inf}\simeq10^{11}\;\rm GeV$, so the amplitude of the tensor power spectrum \eqref{tensor power spectrum dS} and the tensor-to-scalar ratio yield $\Delta^2_h(k_\ast)\simeq4\times 10^{-16}$ and $r(k_\ast) \simeq  2 \times 10^{-7}$, respectively.
Using Eq. \eqref{eq: upper bound particle species}, we find that this corresponds to a maximum number of particle species of $\mathcal{N}_{\rm sp}^{\rm max}\simeq 3 \times 10^{14}$ (dashed blue line).

Lastly, following Ref. \cite{Dvali:2017eba}, in Figure \ref{fig: Hinf VS N our results and Dvali Nsp=1e30}, we also show the case for the maximum number of gravitationally
coupled species $\mathcal{N}_{\rm sp}^{\rm max}\simeq10^{30}$ suggested in Ref. \cite{Dvali:2007wp}.
This constrains the number of e-folds consistent with the classical description of dS inflation to $N_{\rm obs}\simeq126$, i.e. $k_{\rm end}^{-1}\simeq10^{-29}\;\rm m$. Using our results, this corresponds to $H_{\rm inf}\simeq10^{3}\;\rm GeV$, $\Delta^2_h(k_\ast)\simeq4\times~10^{-32}$ and $r(k_\ast)\simeq 2 \times 10^{-23}$. 

\section{The scalar spectral tilt}
\label{sec:The scalar spectral tilt}

We have seen that we can predict the energy scale of inflation $H_{\mathrm{inf}}$ and the tensor-to-scalar ratio $r$, once we fix how long dS lasts, meaning that we fix $N_{\rm obs}$. Moreover, the dimensionless scalar power spectrum $\Delta^2_\phi(k)$ in Eq. \eqref{eq: dimensionless power spectrum to compute} is found to be exactly scale invariant. We discuss in the following how to account for the red-tilt of the power spectrum.

The argument in Refs. \cite{Gomez:2020xdb,Gomez:2021yhd,Gomez:2021yhd} is that time plays a crucial role in defining the tilt of the scalar power spectrum. They show how considering the role of time on dS leads to a prediction of the tilt. Here, we use a more heuristic argument to incorporate time in our scenario.

In our framework, scalar fluctuations are generated via second-order effects from tensor perturbations. We know that different $k$-modes will exit the causal region at different times and the time $t_k$ of horizon crossing during inflation\footnote{$t_k$ highlights the fact that the time of horizon crossing depends on the comoving wavenumber $k$.} is defined by the condition that $k=a(t_k)H(t_k)=\mathcal{H}(t_k)$, where $\mathcal{H}$ is the conformal Hubble parameter. Thus, $t_k$ is the initial time after which we are on super-horizon scales, and the potential $\phi_2$ is constant. Since large scales exit the causal region before small scales, the latter will remain on super-horizon scales for less time. We can implement this by saying that this time scales as $k_L/k$, where $k_L$ is the wavenumber corresponding to the largest scale crossing the Hubble horizon during inflation, and we take it to be the horizon scale $k_L=k_0\simeq1/4000\;\rm Mpc^{-1}$. According to these heuristic arguments, we would expect the spectral tilt to be\footnote{In a range of values around the pivot scale $k_\ast$, we can parametrize the power spectrum as $\Delta^2_\phi(k)=\Delta^2_\phi(k_\ast)(k/k_\ast)^{n_S-1}$.
The scalar spectral tilt is defined as $$n_s-1\equiv \left.\dfrac{d\ln\Delta^2_\phi(k)}{d\ln k}\right|_{k=aH}\;.$$} 
\begin{align}
    n_s-1 \simeq -\dfrac{k_0}{k}\;.
    \label{eq: spectral tilt}
\end{align}
From the latest Planck constraints \cite{Planck:2018jri,Planck:2018vyg}, we have 
that $n_s=0.9649\pm~0.0042$ at the CMB scale (TT, TE, EE+lowE+lensing), so $n_s-1<0$ (red tilt).
From Eq. \eqref{eq: spectral tilt}, the resulting scalar spectral tilt is $n_s-1\sim -\,0.03$ at the pivot scale $k_\ast=0.01\;\rm Mpc^{-1}$, while approaching scale invariance on smaller scales.

\bibliographystyle{apsrev4-2}
%

\end{document}